\documentclass[12pt]{article}
\usepackage{amsfonts}
\usepackage{amsmath}
\usepackage{amssymb}
\usepackage{amsthm}
\usepackage{accents}
\usepackage{youngtab}
\usepackage{braket}
\usepackage{graphicx}
\usepackage[svgnames]{xcolor}
\usepackage{ytableau,tikz}
\usepackage{here}
\usepackage{comment}
\usepackage{simplewick}
\usepackage{tikz}
\usepackage{feynmf}
\usepackage{chessfss}

\def\ba#1\ea{\begin{align}#1\end{align}}

\newcommand{\nn}{\nonumber}

\newcommand{\cN}{\mathcal{N}}
\newcommand{\cW}{\mathcal{W}}

\newcommand{\lt}{\left}
\newcommand{\rt}{\right}

\newcommand{\SU}{\mathrm{SU}}

\makeatletter
\newcommand{\douwidehat}[2]{%
  \sbox0{$\m@th#1\widehat{\hphantom{#2}}$}%
  \sbox2{$\m@th#1x$}
  \sbox4{$\m@th#1#2$}
  \dimen0=\ht0
  \advance\dimen0 -.8\ht2
  \dimen2=\dp4
  \rlap{%
    \raisebox{\dimexpr\dimen0-\dimen2}{%
      \scalebox{1}[-1]{\box0}%
    }%
  }%
  {#2}%
}
\makeatother

\usepackage[colorlinks=true,linkcolor=black,citecolor=black,urlcolor=black]{hyperref}

\begin{document}

\begin{titlepage}

\begin{flushright}
UT-18-27
\end{flushright}

\begin{center}
{\Large
{\bf 
 On the Correspondence between Surface
 Operators in Argyres-Douglas Theories 
%from Higgsing 
and
%Non-vacuum
 Modules 
of %in 
Chiral Algebra
}
\par}
\end{center}

\vskip 1cm

\begin{center}
{\Large
 Takahiro Nishinaka$^\WhiteRookOnWhite$, Shinya Sasa$^\WhiteKnightOnWhite$, %Akihiro Tsuchiya$^\WhiteBishopOnWhite$,\\ 
Rui-Dong Zhu$^\WhiteKnightOnWhite$}
\end{center}

\vskip 1cm

\begin{center}
$^\WhiteRookOnWhite$ {\large {\it Department of Physical Sciences, College of Science and Engineering\\
Ritsumeikan University, Shiga 525-8577, Japan}}\\
$^\WhiteKnightOnWhite$ {\large {\it Department of Physics, The University of Tokyo\\
Bunkyo-ku, Tokyo 113-8654, Japan}}\\
%$^\WhiteBishopOnWhite$ {\large {\it Kavli Institute for the Physics and Mathematics of the Universe,\\ 
%Todai Institutes for Advanced Study, The University of Tokyo\\
%Kashiwanoha, Kashiwa, Chiba 277-8583, Japan}}
\end{center}

\vskip 3cm

\begin{abstract}
We compute the Schur index of Argyres-Douglas theories of type
 $(A_{N-1},A_{M-1})$ with surface operators inserted, via the Higgsing
 prescription proposed by D.~Gaiotto, L.~Rastelli and
 S.~S.~Razamat. These surface operators are obtained by turning on
 position-dependent vacuum expectation values of operators in a UV
 theory which can flow to the
 Argyres-Douglas theories. %In the language of Schur index, 
%this kind of
% surface operator is characterized by poles 
%of %in 
%the wavefunction of the regular puncture. %,
% and we propose a 
We focus on two series of $(A_{N-1},A_{M-1})$ theories;
 one with ${\rm gcd}(N,M)=1$ and
the other with $M=N(k-1)$ for an integer $k\geq 2$. 
%For these two series of Argyres-Douglas
% theories, o
Our results are identified with the characters of non-vacuum modules of the associated 2d chiral algebras,
  which explicitly confirms a remarkable correspondence recently discovered by C.~Cordova, D.~Gaiotto and S.-H.~Shao.
% spanned by Schur operators, for $(A_{k-1},A_{n-1})$ theories with ${\rm gcd}(k,n)=1$ and $(A_{k-1},A_{k(n-1)-1})$ theories. 
\end{abstract}

\end{titlepage}

\tableofcontents

\parskip=6pt

\section{Introduction}

The 4d/2d duality, established after the groundbreaking work of
\cite{AGT,Wyllard:2009hg, GRRY, Gadde:2011uv}, is
remarkable progress %a great triumph
 in the study of 4d $\cN=2$ 
field %gauge
theories. This states that, for a large class of 4d $\cN=2$
theories called class $\mathcal{S}$ theories \cite{classS,
Gaiotto:2009hg}, 
%For a large class of 4d $\cN=2$ theories called class ${\cal
%S}$ theories \cite{classS, Gaiotto:2009hg}, it can be naturally
% understood that
 physical quantities in 4d %gauge 
theories are encoded in a two-dimensional
 theory on a punctured Riemann surface, $\mathcal{C}$. %, as 
This duality is naturally understood when we recall that 
class $\mathcal{S}$ theories %these theories 
are constructed
%from the compactification of 
by compactifying a 6d $\cN=(2,0)$ theory (with ``gauge 
algebra'' $\mathfrak{g}$ %group $G$
) on 
$\mathcal{C}$.\footnote{To keep the 4d $\mathcal{N}=2$
supersymmetry, a partial topological twist is needed on $\mathcal{C}$.}
%some punctured Riemann surface.
The relevant 2d theory on $\mathcal{C}$ depends on the 4d physical
 quantity we consider.
 In particular, when we consider the Schur index of a 4d $\mathcal{N}=2$
 superconformal field theory (SCFT) of class $\mathcal{S}$,
\begin{align}
 \mathcal{I}(q;{\bf x}) \equiv \text{Tr}_{\mathcal{H}}(-1)^F
 q^{E-R}\prod_{\ell=1}^{\mathrm{rank}\,G_F}x_\ell^{f_\ell}~,
\label{eq:Schur}
\end{align}
 %, the superconformal index, or the $S^3\times S^1$ partition function, of a class ${\cal S}$ theory 
the relevant 2d theory is a topological quantum field
 theory (TQFT) called the {\it  $q$-deformed Yang-Mills theory} on
 $\mathcal{C}$. Indeed, the Schur index of class
 $\mathcal{S}$ theories were shown to be identical to the correlation
 function of the $q$-deformed Yang-Mills theory on $\mathcal{C}$
 \cite{GRRY,Gadde:2011uv} (See also \cite{Beem:2016wfs} for a review). In the above definition of the Schur index, $\mathcal{H}$ is the
 Hilbert space of local operators, $E$ and $R$ are respectively the scaling dimension
 and the $\SU(2)_R$ charge, 
 $G_F$ is the flavor symmetry group, and $f_\ell$ are the flavor
 charges.\footnote{We call the global symmetry that commutes with the 4d
 $\mathcal{N}=2$ superconformal symmetry ``flavor symmetry.''} In this
 paper we encounter cases with $G_F=\SU(N)$, in which case we take
 $f_k$ to be minus the $k$-th Cartan generator of $\SU(N)$. The BPS
 operators giving non-vanishing contributions to the Schur
 index are called ``Schur operators.''

In the above discussion, the 2d theory is
 a TQFT and therefore its correlation function is independent of the complex
 structure of $\mathcal{C}$. This reflects the fact that the 4d
 superconformal index is independent of marginal couplings \cite{Gadde:2009kb}. %can be
 %reproduced from a correlator on the corresponding Riemann surface in a
 %2d TQFT called q-deformed Yang-Mills (q-YM) theory  
Indeed, the complex structure deformations of $\mathcal{C}$ correspond
to marginal deformations of the 4d theory \cite{classS}.\footnote{Note
that an $\mathcal{N}=2$ marginal coupling in theories of class
$\mathcal{S}$ SCFTs are {\it not} necessarily encoded in the
complex structure of $\mathcal{C}$. For examples of marginal coupling
arising from a puncture on $\mathcal{C}$, see \cite{Buican:2014hfa,
Xie:2016uqq, Xie:2017vaf,
Buican:2017fiq, Xie:2017aqx}. A relation between the space of marginal
couplings (i.e., conformal manifold)
and the associated chiral algebra was studied in \cite{Buican:2016arp}.} 
On the other hand, each
puncture 
on $\mathcal{C}$ %in this correspondence 
corresponds to an external state 
%in the q-YM, 
of the 2d TQFT, which is naturally associated with % and thus 
a wavefunction of the form, 
\ba
f^{(i)}_R(q;{\bf x}^{(i)}
%x^{(i)}_k
)~.
\ea
Here 
%$(i)$ denotes the species of 
 $i=1,\cdots,m$ is the label of the puncture,
$R$ stands for certain irreducible representation of 
%$G$
$\mathfrak{g}$, and ${\bf x}^{(i)}\equiv \{x^{(i)}_k\}$ are fugacity
parameters associated to the flavor symmetry introduced by the
puncture.\footnote{Note that the $\{{\bf x}^{(1)},\cdots,{\bf x}^{(m)}\}$ can be
 reorganized into ${\bf x} \equiv \{x_1,\cdots, x_{\mathrm{rank}\,G_F}\}$.}
%, which labels the infinite-dimensional Hilbert space
 %of the 2d theory.
Suppose that $\mathcal{C}$ is a Riemann surface of genus $g$ with $m$
punctures. %Given the wavefunction of $m$ punctures on a genus $g$ Riemann
 %surface,
Then 
the correlation function of the 2d TQFT is given by %the superconformal index of the corresponding 4d theory can be computed in the TQFT framework as 
\ba
{\cal I}(q;
{\bf x} %x^{(i)}_k
)=\sum_R \Big(C_R(q)\Big)^{2g-2+m}\prod_{i=1}^mf^{(i)}_R(q;x^{(i)}_k),
\label{eq:TQFT}
\ea
where $C_R(q)$ is 
the structure constant of the TQFT. The 4d/2d duality implies that this
is identical to the Schur index \eqref{eq:Schur} of the corresponding 4d SCFT.
% obtained by taking a suitable residue
%of the wave function for the full regular puncture \cite{GRR-Higgs}.
%often called the structure constant of the q-YM. 

 %We focus on the limit $t\rightarrow q$,
				%$p\rightarrow 0$, which simplifies the
				%expression a lot. It is called and
				%the corresponding index is often referred to as the Schur
%index. 
The Schur index also plays an important role in another 4d/2d
correspondence discovered in \cite{chiralsymm}. This correspondence
states that, for any 4d $\mathcal{N}=2$ SCFT (even without class
$\mathcal{S}$ description), the space of local
operators contributing to the Schur index is equipped with the structure of a chiral
algebra, or equivalently, a vertex operator algebra (VOA). In this context, the Schur
index is identified with the character of the vacuum
module of the chiral algebra.\footnote{For remarkable
unitary/non-unitary relations in this context, see
\cite{Buican:2017rya}. Note also that the associated chiral algebra is
defined in a two-dimensional plane inside the four-dimensional
spacetime. Therefore this 4d/2d correspondence is different from the
AGT correspondence.} %Interestingly, the Schur index counts the number of operators that form
%a 2d chiral algebra (after proper twisting) on the subspace of, say, 3-4
%plane in the 4d $\cN=2$ theory on $\mathbb{R}^4$ \cite{chiralsymm}, and
%thus it gives the vacuum character of the chiral algebra. 
Furthermore, it has recently been shown that the Schur indices in the
presence of a large class of surface operators are identified with the
characters of some non-vacuum modules of the chiral algebra \cite{CGS-17-1, CGS-17-2}.\footnote{It was also shown in
\cite{CGS-line} that, for some Argyres-Douglas theories, the
Schur index in the presence of line defects are also related to
the characters of non-vacuum modules. Its relation to the surface
operator insertions is discussed in \cite{CGS-17-2}.} %One would then also like to reproduce the characters of
		%the remaining non-vacuum modules in the chiral algebra
		%by inserting some operator at the origin of the 3-4
		%plane in the 4d theory side, to establish this 4d/2d
		%duality. 
%The insertion of line defects and surface defects in this setup
%for some simple theories has been considered in \cite{CGS-line,
%CGS-17-1, CGS-17-2}. The Schur index with line defect gives linear
%combination of characters of various modules in the chiral algebra,
%while it seems that for each type of surface defect inserted, the Schur
%index reproduces the character of a single module. 
In particular in \cite{CGS-17-2}, the Schur indices of free hypermultiplets and the $(A_1,A_3),
(A_1,A_5)$ and $(A_1,A_{2n})$
Argyres-Douglas (AD) theories \cite{Argyres:1995jj, Argyres:1995xn,
Eguchi:1996vu, Cecotti:2010fi, Bonelli:2011aa, Xie:2012hs} for all
positive integers $n$, with surface operator insertions, were computed
and shown to
be identical to the characters of
irreducible modules of the associated chiral algebras.

In this article,
% for simplicity on the chiral algebra side, we consider
 we use the TQFT description to compute the Schur indices with surface operator insertions for two
infinite series of AD theories, generalizing the careful
studies in \cite{CGS-17-2}. % (classified in \cite{Xie,
				% Wang-Xie}) of type $(A_{k-1},A_{n-1})$
				% (with $G=A_{k-1}$),
The first series is the $(A_{N-1},A_{M-1})$ theories for integers
 $N,\,M\geq 2$
 satisfying $\mathrm{gcd}(N,M)=1$, whose chiral algebra was
 conjectured in \cite{Cordova-Shao} to be the vacuum sector of the $\mathcal{W}_N$
 minimal model labeled by $(N,N+M)$.\footnote{For an interesting
 observation on the ODE/IM correspondence for these theories, see \cite{Ito:2017ypt,Ito:2018ypt}.} The second series is the
 $(A_{N-1},A_{N(k-1)-1})$ theories for integers $N,\,k \geq
 2$,\footnote{To be precise, the case of $N=k=2$ is not an AD theory but
 the theory of a free hypermultiplet.} whose
 chiral algebra was conjectured in \cite{Creutzig-17, Beem:2017ooy, Creutzig-18,
 Choi-Nishinaka} based on the closed-form formula for the Schur
 indices (without surface operators) obtained in
 \cite{Buican-Nishinaka15, Buican-Nishinaka17}.\footnote{These chiral
 algebras are deeply related to recent progress in the study of VOAs
 appearing in logarithmic CFTs \cite{triplet-W, logrithmic-minimal, AM,
 Tsuchiya-Wood, CRW}. For the Macdonald index and the small
 $S^1$-reduction of
 the $(A_1, A_{2k-3})$ theories, see \cite{Buican:2015tda, Buican:2015hsa}.} 
Note that the first series for $N=2$ and the second series for $N=2$ and
 $k=3,4$ were already studied in
 \cite{CGS-17-2}. Since
 the structure of the non-vacuum modules is generally very complicated,
 generalizing them to the whole theories in the two series is highly non-trivial. For the
 $(A_{N-1},A_{M-1})$ theories with $\mathrm{gcd}(M,N)=1$, we show that
 the Schur indices with surface operator insertions reproduce
 all and only the characters of the irreducible modules of the 
 $(N,M+N)$ $\mathcal{W}_N$ minimal model. For the $(A_{N-1},A_{N(k-1)-1})$ theories, we
 show a similar statement for all $k\geq 2$ with $N=2$. For $N>2$, we
 give a natural conjecture on modules of the logarithmic $\mathcal{B}(k)_{A_{N-1}}$
 algebra \cite{Creutzig-18}. 

% and we try to give the rule of correspondence between surface defects
% contributing to the Schur index and non-vacuum modules in the dual
% chiral algebra. 
%Two most intriguing subclasses among these theories are: (i) the
 %$(A_{k-1},A_{k(n-1)-1})$ series ($k,\ n\in\mathbb{Z}_+$), to which the
 %TQFT approach was proposed in \cite{Buican-Nishinaka15,
 %Buican-Nishinaka17}. 
%The dual chiral algebra for these theories discussed in
 %\cite{Creutzig-17, Creutzig-18, Choi-Nishinaka} are deeply related to
 %recent progress in the study of vertex operator algebra (VOA)
 %\cite{triplet-W, logrithmic-minimal, AM, Tsuchiya-Wood, CRW} (ii)
 %$(A_{k-1},A_{n-1})$ theories with ${\rm gcd}(k,n)=1$, whose dual
 %chiral algebra is conjectured to be a $\cW_{k+1}$ minimal model
 %labeled by $(k+1,k+n+2)$ \cite{Cordova-Shao}. 
%Minimal models have a great advantage that there are only finite numbers of non-degenerate modules in these models, and thus provide us a good starting point to investigate about the realization of non-vacuum modules in 4d side. 

To evaluate the Schur index in the presence of surface operators, 
we %We
 adopt the 
prescription %method 
proposed in \cite{GRR-Higgs}.
%to obtain surface operators lying along the 1-2 plane and sitting at
%the origin of the 3-4 plane in Argyres-Douglas theories,
% i.e. adding 
This prescription first introduces a UV SCFT of class $\mathcal{S}$ whose
Riemann surface $\mathcal{C}$ is obtained by adding a regular puncture
to the original Riemann surface. We then Higgs the flavor symmetry associated
with the added regular puncture, which triggers an RG-flow going back to the
original SCFT.  %Higgsing the flavor symmetry attached
				  %to that puncture with a
				  %position-depending vacuum expectation
				  %value (VEV). 
The main point of the prescription is that, by turning on a position-dependent vacuum expectation
value (VEV) of the Higgs branch operator in this RG-flow, we can
introduce surface
operators in the IR 4d SCFT. %theory.
From the index viewpoint, this operation %Higgsing procedure 
corresponds to 
%take the residue with respect to a corresponding pole in the index for
%theory with the extra puncture.
taking a particular limit of the flavor fugacity so that the UV index is divergent. This divergence reflects the fact
that the there are decoupled massless Nambu-Goldstone (NG) multiplets
associated with the symmetry breaking. The IR index with the surface
operator inserted is then obtained by removing the contributions
of these decoupled NG multiplets from the UV index.

This article is organized as follows. We provide brief reviews on the
TQFT description and the Higgsing prescription for the Schur index in
Sec.~\ref{sec:review}.  In Sec.~\ref{sec:no-flavor}, we study the
Schur indices of the $(A_{N-1},A_{M-1})$ theories with surface operator
insertions for $M$ and $N$ such that $\mathrm{gcd}(M,N)=1$. We
particularly find that all and only the characters of irreducible
modules of the $(N,M+N)$ $\mathcal{W}_N$ minimal model are reproduced from the
Higgsing. In Sec.~\ref{sec:with-flavor}, we study the Schur indices with
surface operator insertions for the $(A_{N-1},A_{N(k-1)-1})$ theory. For
$N=2$, we find that the Schur indices coincide with the characters of
modules of the logarithmic $B_k$ algebra
\cite{Creutzig-17}, by generalizing the check performed in \cite{CGS-17-2}.
% \ref{s:Higgsing}, on the TQFT way to compute the Schur index in
% \ref{s:TQFT}, and on minimal models in section \ref{s:minimal}. We find a beautiful correspondence between the poles and modules in $\cW_2$ and $\cW_3$ minimal models respectively in section \ref{s:W2-minimal} and \ref{s:W3-minimal}, for $(A_1,A_{2n-3})$ series in \ref{s:Bp}, and we propose a general rule in minimal model series and $(A_{k-1},A_{k(n-1)-1})$ series respectively in section \ref{s:general-minimal} and \ref{s:general-Bp}. 

\section{TQFT description and Higgsing prescription}
\label{sec:review}

We here give a brief review of the Schur index of 4d $\mathcal{N}=2$
SCFTs with and without surface operator insertions. In particular,
we review in Sec.~\ref{s:TQFT} the TQFT description of the Schur index
of theories of class $\mathcal{S}$ \cite{GRRY}.
In Sec.~\ref{s:Higgsing}, we review that the
index in the presence of a surface operator is obtained via the
so-called ``Higgsing prescription'' \cite{GRR-Higgs}. 
 
\subsection{TQFT description for Schur index
% and surface-operator factor
%  in 
%$(A_{N-1},A_{M-1})$ %$(A_1,A_n)$
% AD theory
}\label{s:TQFT}

Let us first review how to compute the Schur index of the $(A_{N-1},
A_{M-1})$ theories from the TQFT
picture. These theories are of class $\mathcal{S}$ \cite{classS, Gaiotto:2009hg}, and
therefore associated with a 6d ``gauge algebra'' $\mathfrak{g}$ and a
punctured Riemann surface $\mathcal{C}$.\footnote{While these theories
have class $\mathcal{S}$ constructions, they were originally constructed
in \cite{Cecotti:2010fi} by using type IIB string theory on Calabi-Yau singularities.} As discussed in \cite{Xie:2012hs}, the
$(A_{N-1}, A_{M-1})$ theories are realized by $\mathfrak{g}=A_{N-1}$ and
$\mathcal{C}$ being a sphere with an irregular puncture. Therefore we
focus on $\mathfrak{g}=A_{N-1}$ in this paper. 
To  %Associated to 
each puncture on $\mathcal{C}$, we assign a wavefunction
$f^{(i)}_R(q, {\bf x}^{(i)}
%\{x_i\}
)$ according to the type of the puncture, where we recall that $R$ is an
irreducible representation of $A_{N-1}$, and ${\bf
x}^{(i)}=(x_1^{(i)},\cdots,x_{N-1}^{(i)})$ stands for the fugacities corresponding to the
flavor symmetry associated with the $i$-th puncture. The Schur index of
the theory is then written as \eqref{eq:TQFT},  %, and the Schur index is given by 
% \ba
% {\cal I}(q,{\bf x})=\sum_R (C_R)^{2g-2+m}\prod_{i=1}^m f^{(i)}_R(q,
% {\bf x}^{(i)}%\{x_i\}
% ),
% \ea
where $g$ is the genus of the Riemann surface, and $m$ is the number of
punctures. 
When $\mathcal{C}$ involves an irregular puncture, $g$ must be zero
for the $U(1)_R$ symmetry of the 4d theory to be preserved \cite{Xie:2012hs}.
The factor $C_R$ is given by 
\ba
(C_R)^{-1}=\frac{\chi_R^{\SU(N)}(q^{\rho})}{\prod_{i=2}^{N}(q^i;q)},
%(C_R)^{-1}=\frac{\chi_R^{\SU(N)}(q^{\rho})}{\prod_{i=1}^r(q^{d_i};q)},
\label{eq:C}
\ea
where %with 
%$r=\mathrm{rank}\, \mathfrak{g} = N-1$, %$r={\rm rank}{G}$ 
%the rank of 6d gauge group $G$,
% $d_i$'s are the degree of Casimirs 
%of $\mathfrak{g}$, %in the gauge group,
$\chi_R^{\SU(N)}(x)$ is the character of the $\SU(N)$
 representation $R$, and $\rho=\frac{1}{2}\sum_{\alpha\in
 \Delta^+}\alpha$ is the Weyl vector of $\SU(N)$.\footnote{Here
 $\Delta_+$ is the set of positive roots.} We use the notation such that
 $\chi_R^{\SU(N)}(x)= \sum_{\mu\in P}m_\lambda(\mu)x^\mu$, where
 $\lambda$ is the highest weight of $R$,
 $P$ is the weight lattice of $\SU(N)$, $m_\lambda(\mu)$ is the
 multiplicity of $\mu$ in the representation $R$, and $x^\mu$ is a formal exponential
 such that $x^{\mu_1}x^{\mu_2} = x^{\mu_1+\mu_2}$. In substituting
 $x=q^{\rho}$ in $\chi_R^{\SU(N)}(x)$, we interpret $(q^\rho)^\mu$ to mean $q^{(\rho,\mu)}$.
%with the substitution $x^{-\omega_i} =
% x_i$.

As mentioned above, the expression for the wave function depends on the
type of the puncture. The wavefunction for a full regular puncture \cite{classS} is given by 
\ba
f^{\mathrm{full}}_R(q,
{\bf x}%\{x^i\}
)=\frac{\chi^{\SU(N)}_R({\bf x})}{(q;q)^{N-1}\prod_{\alpha\in \Delta}(q
x^{\alpha}
%x^\alpha
;q)},
\label{eq:full}
\ea
where $\Delta$ is the set of roots of $\mathfrak{g}= A_{N-1}$, and
$\chi_R^{\SU(N)}({\bf x}) \equiv \chi_R^{\SU(N)}(x)$ with the
identifications $x^{-\omega_i}=x_i$ for fundamental weights $\omega_i$.
Especially the wavefunction for $\mathfrak{g}=A_1$ %$N=2$
%$G=$SU(2)
 is given by 
\ba
f^{\mathrm{full}}_R(q,x)=\frac{\chi^{SU(2)}_R(x)}{(q;q)(qx^2;q)(qx^{-2};q)},
\label{eq:full-SU2}
\ea
%and for the representation of dimension ${\rm dim}R=2n+1$ ($n\in \frac{1}{2}\mathbb{Z}_{\geq0}$) of SU(2),  the character reads 
with
$%\ba
\chi^{SU(2)}_R(x)
%=x^{-2n}+x^{-2n+2}+\dots+x^{2n-2}+x^{2n}
=
(x^{{\rm dim}R}-x^{-{\rm dim}R})/(x-x^{-1})%\frac{x^{{\rm dim}R}-x^{-{\rm dim}R}}{x-x^{-1}}
$, where we write $x_1$ as $x$ for simplicity. %\ea
%Note that 
%%there are 
% this wave function has poles at $x^\pm =q^{\frac{s+1}{2}}$ 
% for a non-negative integer $s$, %($s\in \mathbb{Z}$)
% whose %the regular puncture, and the 
% residue %at this pole can be computed to 
% is evaluated as
% \ba
% 2(q;q)^2C_R\Res_{x\rightarrow q^{\frac{s+1}{2}}}x^{-1}f_R(q,x)=(-1)^s q^{\frac{(s+1)^2+1}{2}}q^{-\frac{s}{2}{\rm dim}R}\sum_{k=0}^s q^{k{\rm dim}R}.\label{A1-surface}
% \ea
% This will be important in the discussion of the Higgsing in the next subsection.
% %We may further multiply an overall factor depending on $s$ so that the index at the pole starts from $1$. 

%As for Argyres-Douglas theories, 
Since the Riemann surface $\mathcal{C}$ for AD theories involves an
irregular puncture \cite{Bonelli:2011aa, Xie:2012hs},
we also need %to further specify 
an expression for the wavefunction of the irregular puncture
\cite{Buican-Nishinaka15, Song-TQFT, Buican-Nishinaka17}. 
Indeed, the $(A_{N-1}, A_{M-1})$ theory is realized by $\mathcal{C}$
being a sphere with one irregular puncture. Therefore its Schur index is
written as
\begin{align}
 \mathcal{I}_{(A_{N-1},A_{M-1})}(q,{\bf x}) = \sum_{R}(C_{R})^{-1}\widetilde{f}^{(A_{N-1},A_{M-1})}_R(q;{\bf x})~,
\end{align}
where $R$ runs over all irreducible representations of $A_{N-1}$, and $\widetilde{f}_R^{(A_{N-1},A_{M-1})}$ stands for the wave function for the irregular
puncture. For the $(A_{N-1},A_{M-1})$ theory with $\mathrm{gcd}(N,M)=1$,
the expression for $\widetilde{f}_R$ is simplified since no flavor
symmetry arises from the puncture. Indeed, it is conjectured in
\cite{Song-TQFT} to be given by\footnote{Note here that there is no
flavor fugacity ${\bf x}$ is associated with the irregular puncture in the
case of $\mathrm{gcd}(M,N)=1$.}
\begin{align}
\tilde{f}^{(A_{N-1},A_{M-1})}_R(q) = \oint
 \left(\prod_{k=1}^{N-1}\frac{dz_k}{2\pi i z_k}\right)\Delta({\bf z})\,
 P.E.\left[-\frac{q^{N+M}}{1-q^{N+M}}\chi_{adj}^{SU(N)}({\bf
 z})\right]\chi_R({\bf z})\qquad (\text{if } \mathrm{gcd}(M,N)=1)~,
\label{eq:f-song}
\end{align}
where $\Delta({\bf z}) = \frac{1}{N!}\prod_{1\leq i\leq j\leq N-1}(1-z^{\alpha_i+\cdots+\alpha_j})(1-z^{-(\alpha_i+\cdots+\alpha_j)})$, and
$P.E.\left[g(q,z_i)\right]\equiv \exp\left(\sum_{k=1}^\infty\frac{
g(q^k,z_i^k)}{k}\right)$ for any function $g$ of the fugacities. For
$N=2$ and $M=2n+1$, this expression reduces to
\ba
\widetilde{f}^{(A_1,A_{2n})}_R(q)=\lt\{\begin{array}{cc}
(-1)^mq^{m(m+1)(n+\frac{3}{2})} & {\rm dim}R=2m+1:\ {\rm odd}\\
0 & {\rm otherwise}
\end{array}\rt.~.
\label{eq:f-song-SU2}
\ea

On the other hand, for $(A_{N-1}, A_{M-1})$ theories with
$\mathrm{gcd}(N,M)\neq 1$, the irregular puncture is associated with a
non-trivial flavor symmetry. In particular, when $M=N(k-1)$ for an
integer $k\geq 2$, the irregular puncture is associated with a $U(1)^{N-1}$
flavor symmetry. The expression for the wave function in this case was
conjectured in \cite{Buican-Nishinaka17} as\footnote{For
$(A_1,A_{2k-3})$ theories, this wave
function was first conjectured in \cite{Buican-Nishinaka15}.}
\begin{align}
 \widetilde{f}^{(A_{N-1},A_{N(k-1)-1})}_R(q,{\bf x})=\prod_{j=1}^\infty
 (1-q^j)^{1-N}q^{kC_2(R)}\sum_{
\mu %w
\in R}q^{-\frac{k}{2}F^{ij}
\mu_i\mu_j%w_iw_j
}x^\mu,%x^w
\end{align}
where $\mu$ runs over all weights belonging to the representation $R$,
$(\mu_1,\cdots,\mu_{N-1})$ are the Dynkin labels of $\mu$,
and
$x^{\mu}$ is again the formal exponential such that $x^{\mu} x^{\nu} =
x^{\mu+\nu}$ and $x^{-\omega_i} = x_i$ for the fundamental weights $w_i$.\footnote{Note
that this choice of $x_i$ is slightly different from the one in
\cite{Buican-Nishinaka17}.} The factor $C_2(R)$
%=\frac{1}{2}\langle \lambda(R),\lambda(R)+2\rho\rangle$
 is
the quadratic Casimir invariant, and $(F^{ij})$ is the inverse Cartan matrix:
\begin{align}
(F^{ij})=\frac{1}{N}\lt(\begin{array}{ccccc}
N-1 & N-2 & N-3 & \dots & 1\\
N-2 & 2(N-2) & 2(N-3) & \dots & 2\\
N-3 & 2(N-3) & 3(N-3) & \dots & 3\\
\vdots & \vdots & \vdots & \ddots & \vdots\\
1 & 2 & 3 & \dots & N-1\\
\end{array}\rt).
\end{align}
% The wavefunction for the irregular puncture of type $I_{k,k(n-1)}$, which is used to construct another class of Argyres-Douglas theories $(A_{k-1},A_{k(n-1)-1})$, is given by \cite{Buican-Nishinaka17}
% \ba
% \tilde{f}^{(k,n)}_R(q,x)=\prod_{j=1}^\infty (1-q^j)^{1-k}q^{nC_2(R)}\sum_{w\in R}q^{-\frac{n}{2}F^{ij}w_iw_j}x^w,
% \ea
% where 
% \ba
% (F^{ij})=\frac{1}{k}\lt(\begin{array}{ccccc}
% k-1 & k-2 & k-3 & \dots & 1\\
% k-2 & 2(k-2) & 2(k-3) & \dots & 2\\
% k-3 & 2(k-3) & 3(k-3) & \dots & 3\\
% \vdots & \vdots & \vdots & \ddots & \vdots\\
% 1 & 2 & 3 & \dots & k-1\\
% \end{array}\rt),
% \ea
% and the quadratic Casimir can be evaluated as 
% \ba
% C_2(R)=\frac{1}{2}\langle \lambda(R),\lambda(R)+2\rho\rangle. 
% \ea
% One can check that by taking $k=2$, we recover the wavefunction $\tilde{f}^{(n)}_R(q;x)$ from the above expression. 
In particular, for $N=2$, this expression reduces to 
\ba
\tilde{f}^{(A_1,A_{2k-3})}_R(q;x)=\frac{q^{k\frac{({\rm dim}R)^2-1}{4}}}{(q;q)}{\rm tr}_R\lt[x^{2J_3}q^{-k(J_3)^2}\rt],
\ea
where $J_3$ is the Cartan of $A_1$ such that the fundamental
representation has eigenvalues $J_3 = \pm\frac{1}{2}$.

 %In this article, we first focus on the minimal model series, i.e. $(A_{k-1},A_{n-1})$ with ${\rm gcd}(k,n)=1$, and then go to examine the $(A_{k-1},A_{k(n-1)-1})$ class. 

\subsection{Higgsing prescription}\label{s:Higgsing}

In \cite{GRR-Higgs}, a prescription to evaluate the Schur index in the
presence of a class of surface operators was proposed, which we call the
``Higgsing prescription.''\footnote{See
\cite{Alday:2013kda, Bullimore:2014nla, Chen:2014rca, Maruyoshi:2016caf,  Ito:2016fpl, Nazzal:2018brc, Razamat:2018zel} for earlier applications of this prescription.}  %an RG flow was
				%considered between two SCFTs, where the
				%UV theory is obtained by cutting down a
				%link in the quiver structure of the IR
				%theory and inserting an extra node. 
We here review and describe it in a more
general setup so that we can apply it to the
$(A_{N-1}, A_{M-1})$ theories in the following sections.
Suppose that we are interested in a class
$\mathcal{S}$ theory associated with the Riemann surface
$\mathcal{C}_{\mathrm{IR}}$. To introduce a surface operator in it, we first consider a UV $\mathcal{N}=2$ SCFT associated with a Riemann surface
$\mathcal{C}_{\mathrm{UV}}$ which is obtained by adding an extra
regular puncture to $\mathcal{C}_{\mathrm{IR}}$. 
The original SCFT is recovered by removing the regular puncture from
$\mathcal{C}_{\mathrm{UV}}$, which corresponds to giving a non-vanishing
VEV to 
%the flavor moment map operators 
the Higgs branch operator associated with the
regular puncture. The non-vanishing VEV triggers an RG-flow from
the UV to the IR SCFT. As discussed in \cite{GRR-Higgs}, the existence
of this RG-flow
implies that the index of the IR theory is obtained by taking a limit of
the UV index.\footnote{For another interesting class of RG-flows from AD theories of
class $\mathcal{S}$, see \cite{Buican:2018ddk}.}

To describe this limit, let us focus on the case in which
the additional regular puncture on $\mathcal{C}_\mathrm{UV}$ is a full puncture and therefore associated with an 
$\SU(N)$ flavor symmetry.\footnote{While this Higgsing is possible for
any type of regular puncture, we will only use the full puncture in
this paper.} The original discussion of \cite{GRR-Higgs} focuses on
Higgsing the simple regular puncture, whose generalization to the full regular puncture is
straightforward but needs a little care about the decoupled
Nambu-Goldstone multiplets. See discussions around Eq.(2.44) and in
Sec.~4.3 of \cite{Beem:2014rza} for this generalization without surface
operator insertions.
The existence of the extra full puncture on $\mathcal{C}_\mathrm{UV}$
implies that the UV
SCFT has an additional flavor $\SU(N)$ current. In the superconformal
multiplet containing the flavor current, there exists a scalar Schur operator called a
``flavor moment map'' in the adjoint
representation of the flavor symmetry. We denote by $\mathcal{O}_{\alpha_\ell}$ the flavor moment map
associated with the $\ell$-th simple root, $\alpha_\ell$, of
$\SU(N)$. The contribution from $\mathcal{O}_{\alpha_\ell}$ to the factor $\prod_{k=1}^{N-1}y_k^{f_k}$ in the
Schur index \eqref{eq:Schur} is $y^{\alpha_1}$.\footnote{Here $y^\lambda$ is a
formal exponential such that $y^{-\omega_i} = y_i$ for fundamental
weights $\omega_i$. As mentioned in Sec.~1,
we take the $k$-th flavor charge $f_k$ to be minus the $k$-th Cartan
generator of $\SU(N)$.} This and the fact that every flavor moment map
has $E=2R=1$ implies that
$\mathcal{O}_{\alpha_\ell}$ contributes
$qy^{\alpha_\ell}$ to the Schur index.

To remove the full puncture from the Riemann surface, we need to completely Higgs the flavor
$\SU(N)$ symmetry. This can be achieved by giving a
non-vanishing VEV to $\mathcal{O}_{\alpha_\ell}$ for all
$\ell=1,\cdots,N-1$.\footnote{From the chiral algebra point of view,
this corresponds to the quantum Drinfeld-Sokolov reduction associated
with the principal embedding of $\mathfrak{su}(2)$ in $\mathfrak{su}(N)$.} Let us first focus on  $\mathcal{O}_{\alpha_1}$
associated with the first simple root.
According to \cite{GRR-Higgs}, giving a non-vanishing VEV to
$\mathcal{O}_{\alpha_1}$ triggers an RG-flow whose IR fixed-point has the
following Schur index
\begin{align}
 \mathcal{I}_{\mathrm{vec}}(q)\cdot \lim_{y_1 \to y_1^*}
 \bigg((1-qy^{\alpha_1})\mathcal{I}_{\mathrm{UV}}(q;{\bf x},{\bf y})\bigg)~,
\label{eq:Higgs1}
\end{align}
where $\mathcal{I}_{\mathrm{vec}}(q)\equiv (q;q)^2$ is the Schur index of a
free vector multiplet, and $y_1=y_1^*$ is the solution to the equation
\begin{align}
 qy^{\alpha_1}=1~. 
\end{align}
Note that, since $y^{\alpha_1} = y_1^{-2}y_2$, the above equation
can be solved for $y_1$.
In the expression \eqref{eq:Higgs1}, the factor $(1-qy^{\alpha_1})$ removes the
index contribution of operators
$(\mathcal{O}_{\alpha_1})^n$ for all $n\in\mathbb{N}$. Indeed, these
operators induce a factor $1/(1-qy^{\alpha_1})$ in $\mathcal{I}_{\mathrm{UV}}(q;{\bf
x},{\bf y})$, which is divergent in the limit $y_1\to y_1^*$. This
divergence reflects the fact that there is a massless mode
corresponding to the VEV of
$\mathcal{O}_{\alpha_1}$. A remarkable observation of \cite{GRR-Higgs} is that this
divergent factor, multiplied by $\mathcal{I}_{\mathrm{vec}}(q)^{-1}$, is
the index contribution from the decoupled Nambu-Goldstone (NG) multiplet associated
with the symmetry breaking. The factor
$\mathcal{I}_{\mathrm{vec}}(q)(1-qy^{\alpha_1})$
precisely removes this index contribution, leaving
the index of the IR theory without the decoupled NG sector. 

 While the
index \eqref{eq:Higgs1} still depends on $y_2,\cdots,y_{N-2}$ and $y_{N-1}$, turning on the
VEVs of all  $\mathcal{O}_{\alpha_\ell}$ leads to an
index independent of ${\bf y}$, which is identified with the Schur index
of the IR SCFT associated with the Riemann surface
$\mathcal{C}_{\mathrm{IR}}$. In other words, the IR index is evaluated as
\begin{align}
\mathcal{I}_{\mathrm{IR}}(q;{\bf x}) = 
 \Big(\mathcal{I}_{\mathrm{vec}}(q)\Big)^{\frac{N(N-1)}{2}}\cdot \lim_{{\bf
 y} \to {\bf y}^*}\left(\mathcal{I}_{\mathrm{UV}}(q;{\bf x},{\bf
 y}) \prod_{1\leq i\leq j\leq N-1}\left(1-\prod_{\ell=i}^j(qy^{\alpha_\ell})\right)\right)~,
\label{eq:Higgsing}
\end{align}
where ${\bf y}={\bf y}^*$ is the simultaneous solution to the equations
$qy^{\alpha_\ell} =1$ for $\ell=1,\cdots,N-1$,
or equivalently
\begin{align}
 \prod_{j=1}^{N-1}(y_j)^{c_{\ell j}} =q \qquad \text{for}\qquad \ell=1,\cdots,N-1~,
\end{align}
where $(c_{ij})$ is the Cartan matrix of $\SU(N)$.
Note that turning on the VEV of all $\mathcal{O}_{\alpha_\ell}$ gives rise
to $\frac{N(N-1)}{2}$ decoupled NG multiplets, corresponding to the
positive roots of $\SU(N)$.\footnote{This can be seen explicitly in the wave
function for the full puncture shown in \eqref{eq:full}.} The factor
$(\mathcal{I}_{\mathrm{vec}}(q))^{\frac{N(N-1)}{2}}\prod_{1\leq i\leq
j\leq N-1}(1-\prod_{\ell=i}^j(qy^{\alpha_\ell}))$ then removes the index
contribution from these decoupled NG multiplets.\footnote{As shown in Sec.~4.3 of
\cite{Beem:2014rza}, this result can be regarded as a consequence of
the quantum Drinfeld-Sokolov reduction for the associated chiral algebra.}

Note that
the limit \eqref{eq:Higgs1} can be regarded as taking the residue of the
UV index at $y_1 = y_1^*$. Indeed, this residue computation is the
original form of the prescription proposed in \cite{GRR-Higgs}, in which
 Higgsing flavor $U(1)$ and $\SU(2)$
symmetries is mainly studied. On the other hand, the IR index  \eqref{eq:Higgsing}
obtained by completely Higgsing the $\SU(N)$ flavor symmetry is {\it not} regarded
as the residue of the UV index at ${\bf y}={\bf y}^*$. Indeed, the
residue is the coefficient of
$\prod_{\ell=1}^{N-1}(1-qy^{\alpha_\ell})^{-1}$ in the Laurent series
around ${\bf y}={\bf y}^*$ while \eqref{eq:Higgsing} is essentially
 the coefficient of $\prod_{1\leq i\leq j\leq
 N-1}(1-\prod_{\ell=i}^jqy^{\alpha_{\ell}})^{-1}$. This discrepancy
 arises since the spontaneously broken $\SU(N)$ symmetry implies
$\frac{N(N-1)}{2}$ decoupled NG multiplets corresponding to the number
of positive roots. This point was essentially noted in Sec.~4.3 of \cite{Beem:2014rza}.

%The UV theory has an additional U(1)$_f$ flavor symmetry associated to
%the bifundamental hypermultiplet corresponding to the the inserted
%node, and it is this flavor symmetry in the UV theory that allows us to
%introduce surface defects into the IR theory through the RG flow.
% There are two equivalent ways given in \cite{GRR-Higgs} to add a
% surface operator at the origin of the IR theory. One way is to gauge
% the U(1)$_f$ flavor symmetry and put a vortex defect with respect to
% this new gauge symmetry and then flow to the IR theory. Another one is
% to Higgs the U(1)$_f$ symmetry with a position-dependent VEV. 

Let us now turn to the Schur index with surface operator insertions. For
 $\mathcal{O}_{\alpha_\ell}$ and any positive integer $s_{\ell}$, the Schur
 index also has a contribution from the
derivative operator
$(\sigma^\mu_{+\dot+}\partial_\mu)^{s_\ell}
 \mathcal{O}_{\alpha_\ell}$. Since a single derivative increases the scaling
 dimension by one without changing the $\SU(2)_R$ and flavor
 charges, the index contribution from the derivative operator is
$q^{1+s_{\ell}}y^{\alpha_\ell}$. The UV index
then has a factor $1/(1-q^{1+s_\ell}y^{\alpha_\ell})$, which is divergent at
$q^{1+s_\ell}y^{\alpha_\ell}=1$, or equivalently at
\begin{align}
 \prod_{j=1}^{N-1}(y_j)^{c_{\ell j}} = q^{n_\ell}\qquad
 \text{for}\qquad \ell=1,\cdots,N-1~,
\label{eq:pole-surface}
\end{align}
where we defined ${\bf n}= (n_1,\cdots,n_{N-1})$ for later use by $n_\ell \equiv 1 + s_\ell$.
Let us denote by ${\bf y}={\bf y}^*_{\bf n}$ the simultaneous solution
to the equations \eqref{eq:pole-surface}. 
An important observation of \cite{GRR-Higgs} is that the limit ${\bf
y}\to {\bf y}^*_{\bf n}$ corresponds to turning on a position-dependent
VEV of $\mathcal{O}_{\alpha_\ell}$ and therefore leads to the insertion of a
surface operator labeled by ${\bf n}$ in the IR theory. We denote this surface operator
by $\mathbb{S}^{\bf n}$. According to \cite{GRR-Higgs}, the IR Schur index in the presence of
$\mathbb{S}^{\bf n}$ is evaluated as
\begin{align}
 \mathcal{I}_{\mathrm{IR}}^{\mathbb{S}^{\bf n}}(q;{\bf x}) =
 \mathcal{N}_{\bf n}(q)\,\Big(\mathcal{I}_{\mathrm{vec}}(q)\Big)^{\frac{N(N-1)}{2}}\cdot
 \lim_{{\bf y} \to{\bf y}_{\bf n}^*}\left(
 \mathcal{I}_{\mathrm{UV}}(q;{\bf x},{\bf y})\prod_{1\leq i\leq j\leq
 N-1}\left(1-\prod_{\ell=i}^j(q^{n_\ell}y^{\alpha_\ell})\right)\right)~,
\label{eq:Higgsing2}
\end{align}
where $\mathcal{N}_{\bf n}(q)$ is a factor of the form $aq^{b}$ for
 $a,b\in\mathbb{R}$ so that the $q$-expansion of the index starts with $1$.
 Note here that ${\bf y}^*_{(1,1,\cdots,1)} = {\bf y}^*$, and therefore ${\bf n}=(1,\cdots,1)$
corresponds to the case without surface operators.

In Sec.~\ref{sec:no-flavor}, we study the Schur index with surface
operator insertions for the
$(A_{N-1},A_{M-1})$ theories with
$\mathrm{gcd}(N,M)=1$, via the Higgsing prescription.
In Sec.~\ref{sec:with-flavor}, we turn to the
$(A_{N-1},A_{N(k-1)-1})$ theories.

\section{$(A_{N-1},A_{M-1})$ theories with $\mathrm{gcd}(N,M)=1$}
\label{sec:no-flavor}

In this section, we study the Schur indices of the $(A_{N-1},A_{M-1})$
theories with $\mathrm{gcd}(N,M)=1$ in the presence of various surface
operators. The case of $N=2$ was carefully studied in \cite{CGS-17-2}.

\subsection{Modules in $\mathcal{W}_N$ minimal models}\label{s:minimal}

Since the chiral algebra associated with the $(A_{N-1},A_{M-1})$ theory
is conjectured in \cite{Cordova-Shao} to be the $(N,M+N)$ $\mathcal{W}_{N}$ minimal model, we
here review the
modules in the $\mathcal{W}_N$ minimal models. The $\mathcal{W}_N$ Minimal models, dating back to the celebrated work of
\cite{Fateev-Lukyanov}, are a special class of chiral algebras, which
have only finite number of non-degenerate modules.
%, in contrast to the Verma representation labeled by continuous
%parameters such as the conformal weight of Virasoro algebra. 
%It is also for this reason that makes them the simplest objects to investigate on the non-vacuum modules. 
Interestingly, the vacuum modules of these chiral algebras correspond to
 the $(A_{N-1},A_{M-1})$ AD theories for $\mathrm{gcd}(N,M)=1$
 \cite{Cordova-Shao}. Since these chiral algebras are relatively simple,
 they are a good starting point to study the relation between the
 non-vacuum modules and the surface operators in four dimensions. We
 here give a brief review of the irreducible modules of the $\mathcal{W}_N$
 minimal model.

The $\mathcal{W}_N$ minimal models are usually labeled by three positive integer
parameters, $N$, $P$ and $Q$ (with $P<Q$ and ${\rm gcd}(P,Q)=1$). $N$
denotes the rank of the global part of the underlying
$\cW$-algebra. Each module of the model is further labeled by two
$N$-dim positive-integer-valued vectors, $\vec{n}=(n_i)_{i=1,\dots, N}$
and $\vec{n}'=(n'_i)_{i=1,\dots, N}$, with 
the constraints %one constraint put on each of them 
\ba
\sum_{i=1}^N n_i=Q,\qquad \sum_{i=1}^N n'_i=P.
\ea
We focus on the special case $P=N$, in which case %In the special cases, which we will focus on in the following
%of the paper, $P=N$, 
all $n'_i$'s are forced to take the value $n'_i=1$. Therefore, the
modules
we are interested in %in this class of minimal models
 are parameterized by $N$ positive integers summed to be $Q$. Sometimes,
 it is also convenient to write $Q=N+M$. %, and it
It is well-known that there is a level-rank duality \cite{level-rank} between a rank-$N$ minimal model labeled by $(P=N,Q=N+M)$ and a rank-$M$ minimal model labeled by $(P=M,Q=N+M)$, as one can see from the expression of minimal model central charge, 
\ba
c=(N-1)\lt(1-\frac{(P-Q)^2}{PQ}N(N+1)\rt).
\ea
The conformal weight of the module labeled by $\vec n$ %$\vec{n}$ 
is given by 
%(we further define $\tilde{n}_i:=n_i-1$, which is a non-negative integer and $\sum_i \tilde{n}_i=Q-N=M$), 
\ba
h(
\vec{n}
)=\frac{1}{2(N+M)}\lt(\sum_{i=1}^{N-1}i(N-i)(s^2_i-Ms_i)+2\sum_{i<j}^{N-1} i(N-j)s_is_j\rt)~,
\ea
where we used the notation introduced in the previous section $s_i=n_i-1$.
Note that the above expression
%, even though does not appear to be so, 
is invariant under the permutations among $\tilde{n}_i$'s. Therefore,
independent modules are labeled by $\vec n$ %$\vec{n}$ 
%restricted to 
such that $n_i\leq n_{i+1}$ for $i=1,\dots, N-1$. 
%
% The simplest case in the minimal model series is specified by $(P,Q)=(2,5)$, which is sometimes also called the Lee-Yang singularity model. The central charge of this model is $c=-\frac{22}{5}$, and there are two non-degenerate primary operators with 
% \ba
% h(1,4)=0,
% \ea
% which is the identity operator, and 
% \ba
% h(2,3)=-\frac{1}{5}.
% \ea
%
%Another interesting example is 
% the $(P,Q)=(3,7)$ model with 
%$c=-\frac{114}{7}$ and four non-degenerate primary operators, 
%\ba
%h(1,1,5)=0~,\qquad h(1,2,4)=-\frac{3}{7}~,\qquad h(1,3,3)=-\frac{2}{7}~,\qquad h(2,2,3)=-\frac{5}{7}~.
%\ea
The character of %minimal models for 
the module labeled by $\vec n$ %$\vec{n}$ 
is given by 
\ba
\chi^{(N,N+M)}_{
{\vec n} %\vec{n}
}(q)=\frac{(q^{N+M};q^{N+M})^{N-1}\prod_{i,j}(q^{a_{i,j}};q^{N+M})\prod_{i,j}(q^{N+M-a_{i,j}};q^{N+M})}{(q;q)^{N-1}},
\label{eq:character-minimal}
\ea
where we used the notation 
\ba
a_{i,j}=n_{j}+n_{j+1}+\dots+n_{j+i-1},
\ea
for $i,j\in \mathbb{N}$ %and 
such that $i+j\leq N$. %For example, 
%the vacuum character in the Lee-Yang model can be computed to 
%\ba
%\chi_{(1,4)}(q)=\frac{(q^5;q^5)(q;q^5)(q^4;q^5)}{(q;q)}=1+q^2+q^3+q^4+q^5+2q^6+2q^7+3q^8+{\cal O}(q^9),
%\ea
%and for the $h=-\frac{1}{5}$ module, the character is given by 
%\ba
%\chi_{(2,3)}(q)=\frac{(q^5;q^5)(q^2;q^5)(q^3;q^5)}{(q;q)}=1+q+q^2+q^3+2q^4+2q^5+3q^6+3%q^7+4q^8+{\cal O}(q^9).
%\ea

\subsection{Review of %Example: 
$(A_1,A_{2k})$ series}\label{s:W2-minimal}

%As our first example, let us consider 
Let us first review the $(A_1,A_{2k})$ AD theories, whose Schur index
with surface operator insertions has been evaluated in \cite{CGS-17-2}. The
associated %dual
 chiral algebra is the Virasoro minimal model labeled by
 $(P=2,Q=2k+3)$. 
%The Schur index for this class of theories with surface operator
%inserted has already been investigated in \cite{CGS-17-2}, and we
%provide a review on their results in this section. 

The class $\mathcal{S}$ construction of the $(A_1,A_{2k})$ theory
involves a Riemann surface $\mathcal{C}_{\mathrm{IR}}$ which is a sphere
with one irregular puncture. Therefore the TQFT expression for the Schur index of the $(A_1,A_{2k})$ theory is
given by
\begin{align}
 \mathcal{I}_{(A_1,A_{2n})}(q) = \sum_{R}(C_R)^{-1}\tilde{f}_{R}^{(A_1,A_{2k})}(q)~,
\end{align}
where $R$ runs over the irreducible representations of
$\mathfrak{su}(2)$, and the expressions for $(C_R)^{-1}$
and $\widetilde{f}_R^{(A_1,A_{2k})}$ are shown in \eqref{eq:C} and \eqref{eq:f-song-SU2}.
The above expression for the Schur index was shown to be equivalent to the vacuum character of Virasoro
algebra for $P=2$ and $Q=2n+3$ \cite{Cordova-Shao,Song-TQFT,CGS-17-2}.

The Schur index
with surface operator insertions can be evaluated via the Higgsing
prescription reviewed in Sec.~\ref{s:Higgsing}. We first consider a UV
SCFT associated with the Riemann surface
$\mathcal{C}_\mathrm{UV}$ which is a sphere with the same irregular
puncture and an additional (full) regular puncture.\footnote{For $N=2$, the
only regular puncture is the full puncture.} This UV SCFT is
called the $(A_1, D_{2k+3})$ theory, whose Schur index is evaluated as
\cite{Cordova-Shao, Song-TQFT}
\begin{align}
 \mathcal{I}_{(A_1,D_{2k+3})}(q;y) = \sum_R
 \tilde{f}_{R}^{(A_1,A_{2k})}(q)f^\mathrm{full}_{R}(q;y)~,
\end{align}
where $y$ is the fugacity for the flavor $\SU(2)$ symmetry associated
with the regular puncture, and the expression for $f^\mathrm{full}_R(q;y)$ is shown in
\eqref{eq:full-SU2}. According to the Higgsing prescription, the Schur
index of the IR SCFT in the presence of the surface operator, $\mathbb{S}^{(n)}$, is evaluated as
\begin{align}
 \mathcal{I}_{(A_1,A_{2k})}^{\mathbb{S}^{(n)}}(q) =
\mathcal{N}_n(q)\, \mathcal{I}_\mathrm{vec}(q)\cdot \lim_{y\to
 q^{\frac{n}{2}}}\Big((1-q^{n}y^{-2})\,\mathcal{I}_\mathrm{UV}(q;y)\Big)~,
\label{eq:surface-SU2}
\end{align}
where $n$ is the label of the surface operator, and $\mathcal{N}(q)$ is
a factor of the form $aq^b$ for $a,b\in \mathbb{R}$ so that the
$q$-expansion of the index starts with $1$.
Note that the condition \eqref{eq:pole-surface} now reads $y^2 =
q^n$, and therefore is equivalent to $y=q^{\frac{n}{2}}$. The wave
function for the full puncture $f^\mathrm{full}_R(q;y) =
\chi_R^{\SU(2)}(y)/(q;q)(qy^2;q)(qy^{-2};q)$ has a pole at this point,
which is the contribution of the would-be NG multiplet, as reviewed in Sec.~\ref{s:Higgsing}.
Since this pole is canceled by $(1-q^ny^{-2})$ in \eqref{eq:surface-SU2},
the IR index $\mathcal{I}^{\mathbb{S}^{(n)}}_{(A_1,A_{2k})}$ is
well-defined and evaluated as \cite{CGS-17-2}
\begin{align}
 \mathcal{I}^{\mathbb{S}^{(n)}}_{(A_1,A_{2k})}(q) =
 2^{-1}(-1)^{\alpha}q^{\beta}\,\mathcal{N}_n(q)\,
 \chi^{(2,2k+3)}_{(\bar n,2k+3-\bar n)}(q)~,
\end{align}
where $\bar n = n$ mod $(2k+3)$, and $\alpha$ and $\beta$ are
respectively an integer and
a half-integer determined by $n$ and $k$. Since the $q$-series of $\chi_{(\bar n,
2k+3-\bar n)}^{(2,2k+3)}(q)$ starts with $1$, the factor
$\mathcal{N}_n(q)$ is fixed as  $\mathcal{N}_n(q) =
2(-1)^{-\alpha}q^{-\beta}$, which implies
\begin{align}
 \mathcal{I}^{\mathbb{S}^{(n)}}_{(A_1,A_{2k})}(q) =
 \chi^{(2,2k+3)}_{(\bar n,2k+3-\bar n)}(q)~.
\end{align}
This shows that the surface operator $\mathbb{S}^{(n)}$
corresponds to the $(\bar n, Q-\bar n)$ module of the $(2,Q)$ Virasoro
minimal model. In particular, $\mathbb{S}^{(n)}$ and
$\mathbb{S}^{(n+Q)}$ lead to the same Schur index.

\subsection{%Example:
 $(A_{N-1},A_{M-1})$ series with ${\rm gcd}(N,M)=1$}\label{s:W3-minimal}

We now consider the generalization of the above discussion to the
$(A_{N-1},A_{M-1})$ theories for $N$ and $M$ such that
$\mathrm{gcd}(N,M)=1$. We show that the Schur indices with surface
operator insertions reproduce all and only the characters of irreducible
representations of the $\mathcal{W}_N$ minimal model for $(P,Q) = (N,N+M)$.

The $(A_{N-1},A_{M-1})$ theories are of class $\mathcal{S}$ and associated
with $\mathcal{C}_{\mathrm{IR}}$ which is a sphere with one irregular puncture. As reviewed in
Sec.~\ref{s:TQFT}, this class $\mathcal{S}$
description implies that the Schur
index without surface operators is written as
\begin{align}
 \mathcal{I}_{(A_{N-1},A_{M-1})}(q) = \sum_R (C_R)^{-1}\widetilde{f}_{R}^{(A_{N-1},A_{M-1})}(q)~,
\end{align}
where $\widetilde{f}^{(A_{N-1},A_{M-1})}$ is given in
\eqref{eq:f-song} if $\mathrm{gcd}(N,M)=1$. We now consider an UV SCFT associated with
$\mathcal{C}_{\mathrm{UV}}$ which is obtained by adding a full
regular puncture to $\mathcal{C}_{\mathrm{IR}}$. Its Schur index is
given by 
\begin{align}
 \mathcal{I}_{\mathrm{UV}}(q;{\bf y}) = \sum_R \tilde{f}_R^{(A_{N-1},A_{M-1})}(q)
 \, f^{\mathrm{full}}_R(q,{\bf y})~,
\end{align}
where $f^{\mathrm{full}}_R$ is given by \eqref{eq:full}. As shown in
Sec.~4.3 of \cite{Song:2017oew}, this can be rewritten as\footnote{To be
precise, this is written as
$P.E.\left[\frac{q-q^{M+N}}{(1-q)(1-q^{M+N})}\chi_\mathrm{adj}^{\SU(N)}({\bf
y})\right]$ in \cite{Song:2017oew}. This
rewriting can be achieved by using the identity $\sum_R \chi^{\SU(N)}_R({\bf z})\chi^{\SU(N)}_R({\bf
y}) = \Delta({\bf z})^{-1}\delta({\bf z}-{\bf y})$.}
\begin{align}
 \mathcal{I}_{\mathrm{UV}}(q;{\bf y}) =
 P.E.\left[\left(\frac{q}{1-q}-\frac{q^{M+N}}{1-q^{M+N}}\right)\chi_{\mathrm{adj}}^{\SU(N)}({\bf
 y})\right]~.
\label{eq:UV-ANAM}
\end{align}
Note here that, in terms of the formal exponential $y^\lambda$ such that
 $y_i = y^{-\omega_i}$ for the fundamental weights $\omega_i$ of $\SU(N)$, the character
$\chi_{\mathrm{adj}}^{\SU(N)}({\bf y})$ is written as
\begin{align}
 \chi_{\mathrm{adj}}^{\SU(N)}({\bf y}) = N-1 + \sum_{1\leq i\leq j\leq
 N-1} \big(y^{\alpha_i+\alpha_{i+1} + \cdots + \alpha_j} +
 y^{-(\alpha_i+\alpha_{i+1}+\cdots + \alpha_j)}\big)~,
\end{align}
where $\alpha_i$ are the simple roots of $\SU(N)$.
Therefore the UV index \eqref{eq:UV-ANAM} is expressed as
\begin{align}
 \mathcal{I}_{\mathrm{UV}}(q;{\bf y}) =
 \frac{(q^{M+N};q^{M+N})^{N-1}}{(q;q)^{N-1}}\prod_{1\leq i\leq j\leq
 N-1}
 \frac{(q^{M+N}y^{\alpha_i+\alpha_{i+1}+\cdots+\alpha_j};q^{M+N})(q^{M+N}y^{-(\alpha_i+\alpha_{i+1}+\cdots+\alpha_j)};q^{M+N})}{(qy^{\alpha_i
 + \alpha_{i+1}+\cdots+\alpha_j}
 ;q)(qy^{-(\alpha_i+\alpha_{i+1}+\cdots+\alpha_j)};q)}~.
\label{eq:UV-ANAM2}
\end{align}

Let us now evaluate the Schur index of the $(A_{N-1},A_{M-1})$ theory in
the presence of a surface
operator $\mathbb{S}^{\bf n}$. Via the Higgsing prescription,
it is evaluated as
\begin{align}
 \mathcal{I}_{(A_{N-1},A_{M-1})}^{\mathbb{S}^{\bf n}} (q) =
 \left(\mathcal{I}_{\mathrm{vec}}(q)\right)^{\frac{N(N-1)}{2}}\lim_{{\bf
 y}\to{\bf y}^*_{\bf n }} \left(\mathcal{I}_{\mathrm{UV}}(q;{\bf
 y})\prod_{1\leq i\leq j\leq
 N-1}\bigg(1-\prod_{k=i}^j(q^{n_k}y^{\alpha_k})\bigg)\right)~,
\label{eq:Higgs-ANAM}
\end{align}
where ${\bf y}^*_{\bf n}$ is the simultaneous solution to the equations
\eqref{eq:pole-surface}. Recall that the simple roots are related to
the fundamental weights by $\alpha_i = \sum_{j=1}^{N-1}c_{ij}\omega_j$,
where $(c_{ij})$ is the Cartan matrix. 
Therefore, the equations
\eqref{eq:pole-surface} are equivalent to 
\begin{align}
 y^{\alpha_\ell} = q^{-n_\ell}\qquad \text{for}\qquad \ell=1,\cdots,N-1~.
\end{align}
We see that the factor $\prod_{1\leq i\leq j\leq N-1}(qy^{\alpha_i+\cdots+\alpha_j};q)$ in the
denominator of \eqref{eq:UV-ANAM2} contains a divergent factor canceled by
$\prod_{1\leq i\leq j\leq N-1}(1-\prod_{k=1}^j(qy^{\alpha_k}))$ in \eqref{eq:Higgs-ANAM}.
In particular,
\begin{align}
 \lim_{{\bf y}\to {\bf y}^*_{\bf n}} \frac{\prod_{1\leq i\leq j\leq
 N-1}(1-\prod_{k=1}^j(qy^{\alpha_k}))}{\prod_{1\leq i\leq j\leq
 N-1}(qy^{(\alpha_i+\cdots+\alpha_j)};q)} = \prod_{1\leq i\leq j\leq N-1}\frac{(-1)^{n_i+\cdots+n_j-1}q^{\frac{(n_i+\cdots+n_j)(n_i+\cdots+n_j-1)}{2}}}{(q;q)_{n_i+\cdots+n_j-1}(q;q)}~,
\end{align}
where $(x;q)_k \equiv \prod_{i=1}^{k}(1-q^{i-1}x)$.
Therefore, the IR index is evaluated as
\begin{align}
 \mathcal{I}_{(A_{N-1},A_{M-1})}^{\mathbb{S}^{\bf n}}(q) &= \left(\mathcal{I}_{\mathrm{vec}}(q)\right)^{\frac{N(N-1)}{2}}
 \frac{(q^{M+N};q^{M+N})^{N-1}}{(q;q)^{N-1}} \prod_{1\leq i\leq j\leq
 N-1} \Bigg\{ (-1)^{n_i+\cdots+n_j-1}q^{\frac{(n_i+\cdots+n_j)(n_i+\cdots+n_j-1)}{2}}
\nonumber\\[2mm]
&\qquad\qquad \qquad \times
 \frac{(q^{M+N+(n_i+\cdots+n_j)};q^{M+N})(q^{M+N-(n_i+\cdots+n_j)};q^{M+N})}{(q;q)_{n_i+\cdots+n_j-1}(q^{1+n_i
 + n_{i+1}+\cdots+n_j};q)(q;q)}\Bigg\}~.
\end{align}
Using $\mathcal{I}_{\mathrm{vec}}(q)= (q;q)^2$ and the identities
\begin{align}
 (q;q)_{n-1}(q^{n+1};q) = \frac{(q;q)}{(1-q^n)}~,\qquad
 (1-q^n)(q^{M+N+n};q^{M+N}) = (q^n;q^{M+N})~,
\end{align}
we can rewrite this as
\begin{align}
\mathcal{I}_{(A_{N-1},A_{M-1})}^{\mathbb{S}^{\bf n}}(q) &= 
 \frac{(q^{M+N};q^{M+N})^{N-1}}{(q;q)^{N-1}} \prod_{1\leq i\leq j\leq
 N-1} 
(q^{n_i+\cdots+n_j};q^{M+N})(q^{M+N-(n_i+\cdots+n_j)};q^{M+N})
\nonumber\\
&\qquad \times \mathcal{N}_{\bf n}(q) \prod_{1\leq i\leq j\leq N-1}(-1)^{n_i+\cdots+n_j-1}q^{\frac{(n_i+\cdots+n_j)(n_i+\cdots+n_j-1)}{2}}~.
\end{align}
Recall that $\mathcal{N}_{\bf n}(q)$ is a factor which makes the
$q$-expansion of the IR
index start with $1$. In our case, it is fixed as
\begin{align}
 \mathcal{N}_{\bf n}(q)\equiv \prod_{1\leq i\leq j\leq N-1}(-1)^{n_i+\cdots+n_j-1}q^{-\frac{(n_i+\cdots+n_j)(n_i+\cdots+n_j-1)}{2}}~.
\label{eq:Nn}
\end{align}
Then we see that the IR index
$\mathcal{I}_{(A_{N-1},A_{M-1})}^{\mathbb{S}^{\bf n}}(q)$ is precisely
equivalent to the
character of the irreducible $\mathcal{W}_N$ module
\eqref{eq:character-minimal} 
%labeled by 
for $\vec n \equiv
({\bf n},N+M-\sum_{i=1}^{N-1}n_i)$. 
%shown in
%\eqref{eq:character-minimal}!
  Note here that, up to the prefactor
absorbed in $\mathcal{N}_{\bf n}(q)$, the character
\eqref{eq:character-minimal} is invariant under $n_i\rightarrow n_i+N+M$
for $^\forall i$. Therefore, the two surface operators, $\mathbb{S}^{\bf n}$
and $\mathbb{S}^{{\bf n} + (N+M){\bf 1}}$, are related to the same
module of the $\mathcal{W}_N$ minimal model. For $N=2$ this degeneracy was already
noted in \cite{CGS-17-2}, and we here naturally generalize it to the
$\mathcal{W}_N$ minimal model.
%Up to an overall factor, there is a shift symmetry in the character formula (\ref{eq:character-minimal}) $n_i\rightarrow n_i+N+M$ for $^\forall i$. 

\section{$(A_{N-1},A_{N(k-1)-1})$ theories}
\label{sec:with-flavor}

In this section, we study the Schur indices of the
$(A_{N-1},A_{N(k-1)-1})$ theories in the presence of surface
operators. The chiral algebras associated with these theories were
conjectured to be the logarithmic $\mathcal{B}(k)_{A_{N-1}}$ algebras
\cite{Creutzig-18}. The Schur indices with surface operator insertions for $N=2$
and $k=2,3$ and $4$ were
already computed and shown to be identical to the characters of
modules of the associated chiral algebra in
\cite{CGS-17-2}. We here generalize it to the whole $(A_N,A_{N(k-1)-1})$
theories, and particularly show for the
$(A_1,A_{2k-3})$ theories that the indices are identical to the characters of modules of the $\mathcal{B}(k)_{A_1}$ algebra.

\subsection{Logarithmic $\mathcal{B}(k)_{A_{N-1}}$ algebra}

The chiral algebra associated with the $(A_{N-1},A_{N(k-1)-1})$ theory
was conjectured in \cite{Creutzig-18} to be the logarithmic
$\mathcal{B}(k)_{A_{N-1}}$ algebra constructed in the paper. The
$\mathcal{B}(k)_{A_{N-1}}$ is a nice generalization of the $\mathcal{B}_k$
algebra constructed in \cite{CRW}, the latter of which is
conjectured to be isomorphic to the $W^{(2)}_{k-1}$ algebra
\cite{Feigin:2004wb}.\footnote{The author of \cite{Creutzig-18} indeed constructed the $\mathcal{B}(k)_{\mathcal{Q}}$
algebra for a general simply-laced Lie algebra $\mathcal{Q}$. We focus on the case
$\mathcal{Q}=A_{N-1}$ in this paper.} In this sub-section we briefly review the
structure of the
$\mathcal{B}(k)_{A_{N-1}}$ algebra, following \cite{Creutzig-18}. 

The $\mathcal{B}(k)_{A_{N-1}}$ is a vertex operator algebra defined as
the kernel of screening operators in a lattice vertex operator
algebra.\footnote{The author of \cite{Creutzig-18} also
provided a construction of $\mathcal{B}(k)_{\mathcal{Q}}$ in terms of the ``corner
VOA'' studied in \cite{Creutzig:2017uxh,Gaiotto:2017euk}.} Let $\mathcal{F}_\lambda$ be the weight-$\lambda$ Fock module
of the Heisenberg vertex operator algebra of rank $(N-1)$, and $P$ be
the weight lattice of $\SU(N)$. We then define $\mathcal{M}(k,\mu)\equiv
\bigcap_{j=1}^{N-1}{\rm ker}_{{\cal
F}_{\sqrt{k}\mu}}e_0^{-\alpha_j/\sqrt{k}}$ for $k\in \mathbb{N}$ and
$\mu\in P$, where
$e_0^{-\alpha_j/\sqrt{k}}$ is the screening operator associated with the
$j$-th simple root of $\SU(N)$. The $\mathcal{B}(k)_{A_{N-1}}$ algebra is 
decomposed as
%module 
\ba
{\cal B}(k)_{A_{N-1}}\simeq \bigoplus_{\mu\in P} {\cal M}(k,\mu)\otimes {\cal F}_{\sqrt{-k}\mu}~.
\label{eq:tensor_prod}
\ea
The %Its 
character 
of $\mathcal{M}(k,\mu)$
is evaluated as
\ba
\chi\lt({\cal
 M}(k,\mu)\rt)=\frac{q^{\frac{\rho^2}{2k}}}{\eta(q)^{N-1}}\sum_{\lambda\in
 P^+\cap (Q+\mu)} \left(q^{\frac{k}{2}(\lambda+\rho)^2}m_\lambda(\mu)\sum_{w\in W}\epsilon(w)q^{-\langle \rho,w(\lambda+\rho)\rangle}\right)~,
\ea
where $\eta(q)\equiv q^{\frac{1}{24}}\prod_{i=1}^\infty(1-q^i)$, $P^+$ 
%in the above equation 
denotes the set of dominant weights, $Q$ is the root lattice, and $m_\lambda(\mu)$ counts the
multiplicity of weight $\mu$ in the highest weight representation of $\SU(N)$
associated with $\lambda$. %highest weight.
On the other hand, the character of the Fock module is given by
\begin{align}
 \chi(\mathcal{F}_{\sqrt{-k}\mu}) =
 \frac{q^{-\frac{k\mu^2}{2}}x^\mu}{\eta(q)^{N-1}}~.
\label{eq:Fock-character}
\end{align}
The
tensor product decomposition \eqref{eq:tensor_prod} then implies that
the vacuum character of the $\mathcal{B}(k)_{A_{N-1}}$ is given
by
\begin{align}
\chi(\mathcal{B}(k)_{A_{N-1}}) =  \sum_{\mu\in P}\chi(\mathcal{M}(k,\mu))
\, \chi(\mathcal{F}_{\sqrt{-k}\mu})~,
\end{align}
which was proven in \cite{Creutzig-18} to coincide with the Schur index of the
$(A_{N-1},A_{N(k-1)-1})$ theory, up to the normalization.
%*********** In
%particular, the formal exponential $x$ associated with the Heisenberg
%algebra is interested as the fugacity for the flavor $U(1)^{N-1}$
%symmetry of the $(A_{N-1},A_{N(k-1)-1})$ theory.
In the case of $N=2$, a series of $\mathcal{B}(k)_{A_1}$ modules, $W_s$, is constructed in \cite{CRW, Creutzig-17}. These
modules are of the form 
\begin{align}
W_s = \bigoplus_{r\in
\mathbb{Z}}M_{r+1,s}\otimes \mathcal{F}_{\sqrt{-k}r}~,
\label{eq:module-Ws}
\end{align}
in terms of $M_{r,s}\equiv\bigoplus_{\ell=0}^{\infty}L(h_{r+2\ell,s},\,c_k)$ for
$r\geq 1$ and $M_{r,s}\equiv \bigoplus_{\ell=0}^\infty
L(h_{r-1-2\ell,k-s},\,c_k)$ for $r\leq 0$, where
$L(h_{r,s},c_k)$ is the simple highest weight module of Virasoro
algebra at the highest weight $h_{r,s} \equiv ((kr-s)^2-(k-1)^2)/4k$ and
the Virasoro central charge $c_k\equiv 1-6(k-1)^2/k$. The character of
$M_{r+1,s}$ is evaluated in the reference as
\begin{align}
 \chi(M_{r+1,s}) = \left\{
\begin{array}{l}
 \frac{1}{\eta(q)}\sum_{\ell \geq 0}\left(q^{k\left(\frac{r+1}{2} + \ell
					 -\frac{s}{2k}\right)^2} -
 q^{k\left(\frac{r+1}{2} +\ell + \frac{s}{2k}\right)^2}\right)\qquad
  \text{for}\qquad r\geq 0\\[2mm]
 \frac{1}{\eta(q)}\sum_{\ell\leq 0}\left(q^{k\left(\frac{r-1}{2} +\ell +
					     \frac{s}{2k}\right)^2}-q^{k\left(\frac{r+1}{2}+\ell
									-\frac{s}{2k}\right)^2}\right) \qquad \text{for}\qquad
r\leq -1
\end{array}
\right.~.
\label{eq:character-Mrs}
\end{align}

\subsection{Schur indices with surface operator insertions}

We now study the Schur indices of the $(A_{N-1},A_{N(k-1)-1})$ theory
with surface operator insertions. To that end, we first consider the UV
SCFT associated to a sphere $\mathcal{C}_\mathrm{UV}$ with one irregular
puncture and a full regular puncture. We then Higgs the $\SU(N)$ flavor
symmetry associated with the full puncture to recover the
$(A_{N-1},A_{N(k-1)-1})$ theory as the IR SCFT. The Schur index of the
UV SCFT is evaluated via the TQFT description as
\begin{align}
 \mathcal{I}_\mathrm{UV}(q;{\bf x};{\bf y}) &=
 \sum_{R}\widetilde{f}_R^{(A_{N-1},A_{N(k-1)-1})}(q;{\bf
 x})f^\mathrm{full}_R(q;{\bf y})~,
\label{eq:UV-last}
\end{align}
where $R$ runs over the irreducible representations of $\mathfrak{su}(N)$, and ${\bf y}= (y_1,\cdots,y_{N-1})$ is the fugacity for the flavor
$\SU(N)$ symmetry arising from the full puncture.

According to the Higgsing prescription, the IR index in the presence of
a surface operator $\mathbb{S}^{\bf n}$ is evaluated as
\eqref{eq:Higgsing2}, where ${\bf y}={\bf y}^*_{\bf n}$ is again the
solution to \eqref{eq:pole-surface}. To evaluate this limit, we first note that
\begin{align}
&\Big(\mathcal{I}_\mathrm{vec}(q)\Big)^{\frac{N(N-1)}{2}}\lim_{{\bf y}\to {\bf y}^*_{\bf n}}\Bigg(f^{\mathrm{full}}_R(q;{\bf y}) \prod_{1\leq i\leq j\leq
 N-1}\left(1- \prod_{k=i}^j (q^{n_k}y^{\alpha_k})\right)\Bigg) 
\nonumber\\[2mm]
&= \left(\prod_{1\leq i\leq j\leq N-1}(-1)^{n_i+\cdots+n_j-1}q^{\frac{(n_i+\cdots+n_j)(n_i+\cdots+n_j-1)}{2}}\right)
 \frac{q^{(\nu_{\bf n},\rho)}}{(q;q)^{N-1}}\delta(q^{-\nu_{\bf
 n}})\chi_R^{\SU(N)}(q^{-\nu_{\bf n}})~,
\label{eq:limit-last}
\end{align}
where we use the short-hand notations
\begin{align}
 \nu_{\bf n} \equiv \sum_{i=1}^{N-1}n_i \omega_i~,\qquad \delta(x) \equiv x^{\rho}\prod_{\beta\in \Delta_+}(1-x^{-\beta})~,
\end{align}
and recall that in our notation $(q^{\nu})^\beta = q^{(\nu,\beta)}$ and therefore $\chi_R^{\SU(N)}(q^{\nu}) = \sum_{\mu\in
\lambda}m_{\lambda}(\mu)q^{(\nu,\,\mu)}$ with $\lambda$ being the
highest weight of $R$.
%
% Since
% \begin{align}
%  \sum_{\mu\in \lambda}m_\lambda(\mu) y^\mu &=  \sum_{\mu\in
%  \lambda}m_\lambda(\mu) y^{\sum_{i}(\omega_i,\mu)\alpha_i}~,
% \end{align}
% we find
% \begin{align}
%  \sum_{\mu\in \lambda}m_\lambda(\mu) y^\mu\Big|_{{\bf y}\to{\bf
%  y}^*_{\bf n}} &=  \sum_{\mu\in
%  \lambda}m_\lambda(\mu) q^{-\sum_{i}(n_i\omega_i,\mu)} = \sum_{\mu\in
%  \lambda}m_\lambda(\mu)q^{-(\nu_{\bf n},\mu)} =
%  \chi_R^{\SU(N)}(q^{-\nu_{\bf n}})~,
% \end{align}
%
The identity \eqref{eq:limit-last} directly follows from the expression \eqref{eq:full} for
the wave function associated with the full puncture. Substituting it
and \eqref{eq:UV-last} into the equation \eqref{eq:Higgsing2}, we find
that the IR index in the presence of the surface operator
$\mathbb{S}^{\bf n}$ is given by
\begin{align}
\mathcal{I}_\mathrm{(A_{N-1},A_{N(k-1)-1})}^{\mathbb{S}^{\bf n}}(q;{\bf
 x})
%\mathcal{I}_\mathrm{IR}^{\mathbb{S}^{\bf n}}(q;{\bf x}) 
&= \frac{1}{(q;q)^{N-1}}\sum_R\widetilde{f}^{(A_{N-1},A_{N(k-1)-1})}_R(q;{\bf
 x})q^{(\nu_{\bf n},\rho)}\delta(q^{-\nu_{\bf n}})\chi_R^{\SU(N)}(q^{-\nu_{\bf
 n}})~,
\label{eq:surface-last}
\end{align}
where we used the fact that $\mathcal{N}_{\bf n}(q)$ in
\eqref{eq:Higgsing2} is again fixed as in \eqref{eq:Nn}.
% Note that, when ${\bf n}=(1,1,\cdots,1)$, the IR index
% \eqref{eq:surface-last} correctly reproduce the Schur index of the
% $(A_{N-1},A_{N(k-1)-1})$ theory. Indeed, this corresponds to the case
% without the surface operator, in which case the above limit from the UV
% to the IR was essentially discussed in \cite{Creutzig-18}.tj
% On the other hand, for ${\bf n}\neq (1,1,\cdots,1)$, the expression
% \eqref{eq:surface-last} is regarded as the Schur index of the
% $(A_{N-1},A_{N(k-1)-1})$ theory in the presence of the surface operator
% $\mathbb{S}^{\bf n}$.
To relate this to the character of a %irreducible
module of the associated chiral algebra $\mathcal{B}(k)_{A_{N-1}}$, we
use the formulae
\begin{align}
\widetilde{f}_R^{(A_{N-1},A_{N(k-1)-1})}(q;{\bf x}) &=
 \frac{1}{(q;q)^{N-1}}q^{\frac{k}{2}(\lambda,\lambda+2\rho)}\sum_{\mu\in
 Q+\lambda}m_\lambda(\mu)q^{-\frac{k}{2}\mu^2}x^\mu~,
\label{eq:rewriting1}
\\
\delta(q^{-\nu_{\bf n}})\chi_R^{\SU(N)}(q^{-\nu_{\bf n}}) &= \sum_{w\in
 W}\epsilon(w)q^{-(\nu_{\bf n},w(\lambda+\rho))}~,
\label{eq:rewriting2}
\end{align}
where $\lambda$ is the highest weight of $R$, $W$ is the Weyl group of $\mathfrak{su}(N)$, and $\epsilon(w)$
is the sign of the Weyl reflection. The first rewriting
\eqref{eq:rewriting1} was obtained in \cite{Creutzig-18} while the second
one \eqref{eq:rewriting2} follows from the Weyl's character
formula. Substituting these two into \eqref{eq:surface-last} and
performing a small computation as in Sec.~3 of \cite{Creutzig-18}, we obtain
% \begin{align}
%  \mathcal{I}_\mathrm{IR}^{\mathbb{S}^{\bf n}}(q;{\bf x}) &=
%  \frac{1}{(q;q)^{2(N-1)}}\sum_{\lambda\in P^+}\left(
%  q^{\frac{k}{2}(\lambda,\lambda+2\rho)}\sum_{\mu\in Q+\lambda}m_\lambda(\mu)q^{-\frac{k}{2}\mu^2}x^\mu\sum_{w\in W}\epsilon(w)q^{-(\nu_{\bf
%  n},w(\lambda+\rho)+\rho)}\right)~.
% \end{align}
\begin{align}
\mathcal{I}_\mathrm{(A_{N-1},A_{N(k-1)-1})}^{\mathbb{S}^{\bf n}}(q;{\bf x})
% \mathcal{I}_\mathrm{IR}^{\mathbb{S}^{\bf n}}(q;{\bf x}) 
&=
 \frac{q^{-\frac{k}{2}\rho^2-(\nu_{\bf n},\rho)}}{(q;q)^{2(N-1)}}\sum_{\mu\in P}\left(
 \sum_{\lambda\in P^+\cap(Q+\mu)}q^{\frac{k}{2}(\lambda+\rho)^2} m_\lambda(\mu)q^{-\frac{k}{2}\mu^2}x^\mu\sum_{w\in W}\epsilon(w)q^{-(\nu_{\bf
 n},w(\lambda+\rho))}\right)~.
\end{align}
In terms of \eqref{eq:Fock-character} and 
\begin{align}
 \chi(k,\mu,\nu) \equiv \frac{q^{\frac{\nu^2}{2k}}}{\eta(q)^{N-1}}\sum_{\lambda \in P^+\cap
 (Q+\mu)}q^{\frac{k}{2}(\lambda+\rho)^2}m_\lambda(\mu)\sum_{w\in
 W}\epsilon(w)q^{-(\nu,w(\lambda+\rho))}~,
\label{eq:chi-kmn}
\end{align}
we finally have
\begin{align}
\mathcal{I}_\mathrm{(A_{N-1},A_{N(k-1)-1})}^{\mathbb{S}^{\bf n}}(q;{\bf x})
 =
 q^{-\frac{\rho^2+\nu_{\bf n}^2+2k(\nu_{\bf
 n},\rho)}{2k}-\frac{N-1}{12}}\sum_{\mu\in P}\chi(k,\mu,\nu_{\bf
 n})\,\chi(\mathcal{F}_{\sqrt{-k}\mu})~.
\label{eq:final-last}
\end{align}

\subsection{Comparison to the characters of $\mathcal{B}(k)_{A_{N-1}}$ modules}
\label{subsec:comparison-last}

Let us now compare the above result with the character of
modules of the associated chiral algebra $\mathcal{B}(k)_{A_{N-1}}$. We
start with the case of $N=2$, and write $\mu = r\omega_1$ and
$\nu=s\omega_1$ in terms of the fundamental weight $\omega_1$ of
$\SU(2)$. In this case, the expression for $\chi(k,\mu,\nu)$ reduces to
\begin{align}
 \chi(k,\mu,\nu) &=
 \frac{q^{\frac{s^2}{4k}}}{\eta(q)}\sum_{\ell=0}^\infty
 q^{\frac{k}{4}(2\ell + |r|+1)^2}\left(q^{-\frac{s(2\ell+|r|+1)}{2}}-q^{\frac{s(2\ell+|r|+1)}{2}}\right)~.
\end{align}
After a small computation, we see
that this is precisely equivalent to the character of the Virasoro
module $M_{r+1,s}$ shown in \eqref{eq:character-Mrs}, i.e.,\footnote{Note here that,
since it is a module of the Virasoro algebra at
$c=1-6(k-1)^2/k$, the $M_{\mu+1,\nu}$ implicitly depends on $k$.}
\begin{align}
 \chi(k,\mu,\nu) = \chi(M_{r+1,s})~.
\end{align}
This and \eqref{eq:final-last} imply that
\begin{align}
\mathcal{I}_\mathrm{(A_1,A_{2k-3})}^{\mathbb{S}^{(n)}}(q;x) =
 q^{-\frac{1+\nu^2+2k\nu}{4k}-\frac{1}{12}}\sum_{r=-\infty}^{\infty}\chi(M_{r+1,n})\chi(\mathcal{F}_{\sqrt{-k}r})
 = q^{-\frac{1+\nu^2+2k\nu}{4k}-\frac{1}{12}}\chi(W_n)~,
\end{align}
where $W_n$ is the $\mathcal{B}(k)_{A_1}$-module $W_s$ for $s=n$ reviewed in
\eqref{eq:module-Ws}. Hence, up to the prefactor
$q^{-\frac{1+\nu^2+2k\nu}{4k}-\frac{1}{12}}$, the IR index
$\mathcal{I}_\mathrm{(A_1,A_{2k-3})}^{\mathbb{S}^{(n)}}(q;{\bf x})$ in the presence
of the surface operator $\mathbb{S}^{(n)}$ is identical to the character of the
$\mathcal{B}(k)_{A_1}$ module $W_n$. In particular, the label $n$ of the
surface operator is now interpreted as the label of the
$\mathcal{B}(k)_{A_1}$-module.
The different prefactor
arises here since we normalize our IR index so that its $q$-expansion
starts with $1$. This correspondence between the surface operators
$\mathbb{S}^{(n)}$ and the $\mathcal{B}(k)_{A_1}$-modules $W_{n}$ can be
regarded as a generalization of the detailed discussions on $k=2,3$ and $4$
in \cite{CGS-17-2}.

For general $N>2$, modules of the
$\mathcal{B}(k)_{A_{N-1}}$ are not well-studied. However, after the above
success in the $N=2$ case, it is natural to expect that there exists a
higher-rank generalization, $\mathcal{M}(k,\mu,\nu)$, of the module
$M_{r,s}$ so that its character $\chi(\mathcal{M}(k,\mu,\nu))$ coincides
with \eqref{eq:chi-kmn}. Note that $M_{r,s}$ can be realized as the
kernel of a screening charge $\mathcal{Q}_-^{[s]}$ \cite{Tsuchiya-Wood},
 where
$\mathcal{Q}_-^{[s]}$ is the zero mode of the product of $s$ screening
currents acting on the Fock module $\mathcal{F}_{\alpha_{r,s}}$ with
$\alpha_{r,s} \equiv
\frac{1-r}{2}\sqrt{2k}-\frac{1-s}{2}\sqrt{\frac{2}{k}}$. Therefore we
expect that $\mathcal{M}(k,\mu,\nu)$ is also realized in the same way as
the kernel of a screening charge.
We then conjecture that there exists a module $W_{\nu}$ of
$\mathcal{B}(k)_{A_{N-1}}$ which is decomposed as
\begin{align}
 W_\nu = \bigoplus_{\mu\in P}\mathcal{M}(k,\mu,\nu)\otimes \mathcal{F}_{\sqrt{-k}\mu}~.
\end{align}
This means that $\mathcal{I}_\mathrm{(A_{N-1},A_{N(k-1)-1})}^{\mathbb{S}^{\bf n}}(q;{\bf x})$ coincides with the
character of $W_{\nu_{\bf n}}$ up to the prefactor $q^{-\frac{\rho^2+\nu_{\bf n}^2
+ 2k(\nu_{\bf n},\rho)}{2k}-\frac{N-1}{12}}$. We will leave a careful
study of this conjecture for future work.

\section{Conclusion and Discussion}

In this article, we 
%initiated a program to investigate the correspondence between poles in
%the wavefunction of regular puncture and modules in the dual chiral
%algebra.
have studied the Schur indices of two series of $(A_{N-1},A_{M-1})$ Argyres-Douglas
 theories in the presence of surface operators; one for $N$ and $M$ satisfying $\mathrm{gcd}(N,M)=1$ and the other for $M=N(k-1)$. 
% We end up with beautiful rules such as (\ref{rule-minimal}) and
 %(\ref{rule-Bp}) respectively for the $(A_{k-1},A_{n-1})$ series (with
% ${\rm gcd}(k,n)=1$) and the $(A_{k-1}, A_{k(n-1)-1})$ series. 
We have used
 the Higgsing prescription proposed in \cite{GRR-Higgs} and shown,
 for all theories in the first series and theories with $N=2$ in the
 second series, that
 the Schur indices in the presence of the surface operators
 $\mathbb{S}^{\bf n}$ are identical to the characters of non-vacuum
 modules of the associated chiral algebras in the sense of
 \cite{chiralsymm}, which explicitly confirms the recent proposal on the
 general correspondence between surface operators and modules of chiral
 algebras \cite{CGS-17-2}. We have also given a conjecture on
 modules of the $\mathcal{B}(k)_{A_{N-1}}$ algebra based on our results
 on the second series of the Argyres-Douglas theories for $N>2$.

As mentioned in Sec.~\ref{subsec:comparison-last}, %Note that 
non-degenerate modules $M_{r,s}$ in Virasoro minimal models can %also 
be
expressed in terms of the kernel and image of screening charges ${\cal
Q}^{[s]}_-$ (see \cite{Tsuchiya-Wood}). %,
We % we 
expect a more general
corresponding principle exists %to exist 
between poles and screening charges in
the VOA approach. We leave it as a future work together with the
extension to other %gauge 
theories such as the Argyres-Douglas theories of type $(A_{k-1},D_{n-1})$. 

The Macdonald limit 
%$q\neq t$ 
of the superconformal index \cite{Gadde:2011uv}  with surface operator
inserted is also an interesting 
object %subject 
to study, as it gives a refined character 
of the space of %to
the same set of local operators. %-operator excitations.
 As discussed in \cite{chiralsymm, Song-Macdonald, Beem:2017ooy, Bonetti:2018fqz}, the refinement is realized by a
 new grading for the number of ``basic" operators used to produce a
 descendant state, 
which %and
 in Virasoro minimal models
%, it 
counts the number of $L_{-^\forall n}$'s in each state. At least for the
vacuum module, a 
clear %clean
 way to add this new grading in the computation of the character can be
 found in the POSET approach to minimal models \cite{FNMZ, BFS-burge}, 
whose generalization to higer-rank cases and non-vacuum modules is worth
studying. %and we would like to examine how to generalize this prescription into higher rank and non-vacuum module cases. 

Last but not least, a similar Higgsing procedure also exists in the
computation of the instanton partition function (or the $S^4$ partition
function) \cite{Gaiotto-Kim}, as it can be uplift to the superconformal
index of 5d $\cN=1$ theories. In terms of the topological string, this
Higgsing procedure corresponds to the geometric transition and surface
operators obtained this way are the so-called monodromy defects studied
in \cite{Gukov-Witten1,Gukov-Witten2} (see also \cite{Nilpotent-defect}
for an approach from 6d $\cN=(2,0)$ theories) labeled by a nilpotent
elements in $\mathfrak{g}$. When $\mathfrak{g}=\mathfrak{su}(n)$,
nilpotent elements are classified by the partition of $n$, $\{n_i\}$
with $\sum_i n_i=n$. On the other hand, modules in minimal models with
level-rank duality are also labeled by such a partition. 
%Their 
The relation 
between them,
 together with its generalization to the other chiral algebras, 
is an interesting problem to work on.
%, and certainly one wants to know how this kind of label might be realized in other chiral algebras. 

\section*{Acknowledgement}

We would like to thank Tomoyuki Arakawa, Matthew Buican, Thomas Creutzig, Yutaka Matsuo,
Jaewon Song and Akihiro Tsuchiya for helpful
discussion. T.~N. particularly thanks Matthew Buican for various
illuminating discussions in many collaborations on the superconformal indices and chiral algebras of
Argyres-Douglas theories. The work of T.~N. is partially
supported by JSPS KAKENHI Grant Number 18K13547. R.~Z. is supported by JSPS fellowship for young students.

\appendix

\vskip 1cm

\begin{flushleft}
{\LARGE {\bf Appendix}}
\end{flushleft}

\section{Explicit $q$-series for lower-rank examples}

We here give the explicit $q$-series expression for the Schur indices
with surface operator insertions for lower-rank $(A_{N-1},A_{M-1})$
theories with $\mathrm{gcd}(M,N)=1$.
%We remark that all these poles we want to take residue about reside in
%the wavefunction of the full regular puncture, and schematically we have 
In this appendix, we use the fact that the Higgsing prescription gives
rise to the following replacement of
the wave function for the full regular puncture:
\ba
(q;q)^{|\Delta|}f^{\rm full}_R(q,y_i)\rightarrow f_{\vec{t}}(q)\frac{\chi_R(q^{t_i})}{\chi_R(q^\rho)}C^{-1}_R,\qquad y_i\rightarrow q^{t_i},\label{sch-puncture-residue}
\ea
where $|\Delta|=\frac{N(N-1)}{2}$ is the number of roots of $\SU(N)$,
and $t_i$ are determined by ${\bf n}$ so that 
\begin{align}
 n_i = \sum_{j}c_{ij}t_j~,
\end{align}
where $(c_{ij})$ is the Cartan matrix of $\SU(N)$.
This replacement %which 
allows us to interpret the factor
$f_{\vec{t}}(q)\frac{\chi_R(q^{t_i})}{\chi_R(q^\rho)}$ as the
contribution from the surface defect $\mathbb{S}^{{\bf n}}$ to the Schur index. Especially
$f_{\vec{t}}(q)$ does not depend on the representation of $\SU(N)$,
%$G$,
 and when
we take $\{q^t_i\}=q^\rho$, $\mathbb{S}^{{\bf n}}$ becomes a trivial
defect with $f_{\vec{t}}(q)=1$. 

We need the explicit expression of $\tilde{f}^{(A_{N-1},A_{M-1})}_R(q)$ given in \ref{eq:f-song} to perform the $q$-series computation. The wavefunction series $\tilde{f}^{(A_{N-1},A_{M-1})}_R(q)$  can be generated from a formal wavefunction $ \psi_R^{I_{N,-N+1}}(q)$ by 
\ba
\tilde{f}^{(A_{N-1},A_{M-1})}_R(q)= \psi_R^{I_{N,-N+1}}(q^{N+M})~,
\ea
and as shown in \cite{Song-TQFT}, $ \psi_R^{I_{N,-N+1}}(q)$ can be evaluated to 
\ba
\psi^{I_{N,-N+1}}_R(q)= \sum_{n_\alpha \in\mathbb{Z}}\frac{(-1)^{\sum_\alpha n_\alpha}q^{\sum_\alpha \frac{1}{2}n_\alpha(n_\alpha+1)}}{|\cW|(q;q)^{\frac{1}{2}(N-1)(N-2)}}\sum_{w\in{\cal W}}\epsilon(w)\delta_{w\cdot \lambda(R)=\sum_{\alpha\in\Delta^+}n_\alpha \alpha},
\ea
where $w\cdot \lambda=w(\lambda+\rho)-\rho$ is a shifted Weyl reflection, and $\lambda(R)$ stands for the highest weight of the representation $R$ of SU($N$). 
With the help of the (generalized) Euler identities given in Appendix \ref{a:Euler-id}, one can explicitly calculate $\psi^{I_{N,-N+1}}_R(q)$ for low-dimensional representations $R$\footnote{Note that the fundamental weights are expressed in terms of simple roots as $\mu_1=\frac{2}{3}\alpha_1+\frac{1}{3}\alpha_2$, $\mu_2=\frac{1}{3}\alpha_1+\frac{2}{3}\alpha_2$.}. In particular, for $N=3$, the closed form of $\psi^{I_{3,-2}}_{R}(q)$ is conjectured in \cite{Song-TQFT} to be 
\ba
\psi^{I_{3,-2}}_{(\lambda_1,\lambda_2)}(q)=\lt\{\begin{array}{cc}
q^{k(k+1)+l(l+1)+kl} & \lambda_1=3k,\ \lambda_2=3l,\\
-q^{k^2+l^2-1+(k-1)(l-1)} & \lambda_1=3k-2,\ \lambda_2=3l-2,\\
0 & {\rm otherwise},
\end{array}\rt.\label{su3-irreg}
\ea
where $(\lambda_1,\lambda_2)$ stands for the highest weight of an irreducible representation of SU(3). Similarly, one can see that in the case of $N=4$, the first several non-trivial contributions, $\psi_{R}^{I_{4,-3}}(q)$, come from the trivial representation {\bf 1}, the {\bf 15}-dim representation $(1,0,1)$, two {\bf 45}-dim representations $(2,1,0)$ and $(0,1,2)$, two {\bf 35}-dim representations $(4,0,0)$ and $(0,0,4)$ given by\footnote{In SU(4), the fundamental weights are given by $\mu_1=\frac{3}{4}\alpha_1+\frac{1}{2}\alpha_2+\frac{1}{4}\alpha_3$, $\mu_2=\frac{1}{2}\alpha_1+\alpha_2+\frac{1}{2}\alpha_3$, $\mu_3=\frac{1}{4}\alpha_1+\frac{1}{2}\alpha_2+\frac{3}{4}\alpha_3$.}  
\ba
\psi_{(0,0,0)}^{I_{4,-3}}=1,\quad \psi_{(1,0,1)}^{I_{4,-3}}=-q,\quad \psi_{(2,1,0)}^{I_{4,-3}}=\psi_{(0,1,2)}^{I_{4,-3}}=q^2,\quad \psi_{(4,0,0)}^{I_{4,-3}}=\psi_{(0,0,4)}^{I_{4,-3}}=-q^3,\quad\dots
\ea
We note that the contribution from the {\bf 20}-dim representation $(0,2,0)$ also vanishes. 

Based on the conjectured Euler identity for $N=5$ given in Appendix \ref{a:Euler-id}, we also list the first several non-trivial wavefunctions for $\psi^{I_{5,-4}}_R(q)$: 
\ba
\psi_{(0,0,0,0)}^{I_{5,-4}}=1,\quad \psi_{(1,0,0,1)}^{I_{5,-4}}=-q,\quad \psi_{(2,0,1,0)}^{I_{5,-4}}=\psi_{(0,1,0,2)}^{I_{5,-4}}=q^2,\quad \dots
\ea

\subsection{%Example:
 $(A_2,A_{n-1})$ series with ${\rm gcd}(3,n)=1$}\label{a:W3-minimal}

Let us 
first %next
 examine $(A_2,A_{n-1})$ theories with ${\rm gcd}(3,n)=1$. 
We start with $(A_2,A_3)$ theory, whose Schur index is given by 
\ba
{\cal I}_{(A_2,A_3)}&=\frac{1}{(q^2;q)(q^3;q)}\sum_{(\lambda_1,\lambda_2)} \chi_{(\lambda_1,\lambda_2)}(q^\rho)\psi^{I_{3,-2}}_{(\lambda_1,\lambda_2)}(q^7)\nn\\
&=\frac{1}{(q^2;q)(q^3;q)}\lt(1-q^7\sum_{(w_1,w_2)\in R_{(1,1)}}q^{w_1+w_2}+{\cal O}(q^{11})\rt)\nn\\
&=\frac{1}{(q^2;q)(q^3;q)}\lt(1-q^5-2q^6-2q^7-2q^8+{\cal O}(q^{9})\rt)\nn\\
&=1+q^2+2q^3+3q^4+3q^5+6q^6+7q^7+11q^8+{\cal O}(q^9).
\ea
It agrees with the vacuum character of the $(P=3,Q=7)$ minimal model. 

Recall that the wave function of the regular puncture for $G=$SU(3) is given by 
\ba
f^{\rm full}_R(q,y_1,y_2)=\frac{\chi_R(y_1,y_2)}{(q;q)^2(qy^2_1y_2^{-1};q)(qy^{-1}_1y_2^{2};q)(qy_1y_2;q)(qy^{-2}_1y_2^{1};q)(qy^{1}_1y_2^{-2};q)(qy^{-1}_1y^{-1}_2;q)}.
\ea
The insertion of a trivial surface operator corresponds to the reduction\footnote{As explained in section \ref{s:W3-minimal}, we need to compensate a factor ${\cal N}_{\bf n}(q)$ in this calculation so that the IR index starts from $1$. For our convenience, we include this factor implicitly in this kind of reduction denoted by $\rightarrow$.}
\ba
(q;q)^6f^{\rm full}_R(q,y_1,y_2)\rightarrow C^{-1}_R,\quad y_1,y_2\rightarrow q,
\ea
and thus the Schur index with a trivial surface operator $\mathbb{S}^{{\bf n}=(1,1)}$ inserted reproduces the vacuum character. 

The first non-trivial pole, that gives rise to a non-vacuum character, appears at $y_1=q^{\frac{4}{3}}$ and $y_2=q^{\frac{5}{3}}$. Sitting on this pole, the regular puncture reduces to the factor, 
\ba
(q;q)^6f^{\rm full}_R(q,y_1,y_2)\rightarrow \frac{1+q+q^2}{(q^2;q)(q^3;q)}\chi_R(q^{\frac{4}{3}},q^{\frac{5}{3}}),
\ea
and the corresponding Schur index of $(A_2,A_3)$ with this insertion can be computed as 
\ba
{\cal I}_{(A_2,A_3)}^{\mathbb{S}^{(1,2)}}(q)
&=\frac{1+q+q^2}{(q^2;q)(q^3;q)}\lt(1-q^7\sum_{(w_1,w_2)\in R_{(1,1)}}q^{\frac{4}{3}w_1+\frac{5}{3}w_2}+q^{14}\sum_{\substack{(w_1,w_2)\in R_{(3,0)}\\(w_1,w_2)\in R_{(0,3)}}}q^{\frac{4}{3}w_1+\frac{5}{3}w_2}+{\cal O}(q^{20})\rt)\nn\\
&=\frac{1+q+q^2}{(q^2;q)(q^3;q)}\lt(1-q^4-q^5-q^6-2q^7-q^8+{\cal O}(q^{10})\rt)\nn\\
&=1+q+2q^2+3q^3+5q^4+7q^5+11q^6+14q^7+21q^8+28q^9+{\cal O}(q^{10}).
\ea
It agrees with the character of the $(1,2,4)$ module in the $(P=3,Q=7)$ minimal model. Similarly, the pole at $y_1=q^{\frac{5}{3}}$ and $y_2=q^{\frac{7}{3}}$ leads to 
\ba
(q;q)^6f^{\rm full}_R(q,y_1,y_2)\rightarrow \frac{(1+q^2)(1+q+q^2)}{(q^2;q)(q^3;q)}\chi_R(q^{\frac{5}{3}},q^{\frac{7}{3}}),\quad y_1\rightarrow q^{\frac{5}{3}},\ y_2\rightarrow q^{\frac{7}{3}},
\ea
and 
\ba
{\cal I}_{(A_2,A_3)}^{\mathbb{S}^{(1,3)}}(q)&=\frac{(1+q^2)(1+q+q^2)}{(q^2;q)(q^3;q)}\lt(1-q^7\sum_{(w_1,w_2)\in R_{(1,1)}}q^{\frac{5}{3}w_1+\frac{7}{3}w_2}+q^{14}\sum_{\substack{(w_1,w_2)\in R_{(3,0)}\\(w_1,w_2)\in R_{(0,3)}}}q^{\frac{5}{3}w_1+\frac{7}{3}w_2}+{\cal O}(q^{17})\rt)\nn\\
&=\frac{(1+q^2)(1+q+q^2)}{(q^2;q)(q^3;q)}\lt(1-q^3-q^4-q^6-q^7-q^8+q^9+{\cal O}(q^{10})\rt)\nn\\
&=1+q+3q^2+3q^3+6q^4+8q^5+13q^6+17q^7+25q^8+33q^9+{\cal O}(q^{10}).
\ea
It reproduces the character of the $(1,3,3)$ module in the $(P=3,Q=7)$ minimal model. The pole at $x_1=q^{\frac{7}{3}}$ and $x_2=q^{\frac{8}{3}}$ leads to the factor 
\ba
f_{(\frac{7}{3},\frac{8}{3})}(q)=(1+q+q^2)(1+q+q^2+q^3+q^4),
\ea
and the index 
\ba
{\cal I}_{(A_2,A_3)}^{\mathbb{S}^{(2,3)}}(q)
&=\frac{(1+q+q^2)(1+q+q^2+q^3+q^4)}{(q^2;q)(q^3;q)}\lt(1-q^7\sum_{(w_1,w_2)\in R_{(1,1)}}q^{\frac{7}{3}w_1+\frac{8}{3}w_2}\rt.\nn\\
&\qquad\qquad\qquad\qquad\qquad\qquad \lt.+q^{14}\sum_{\substack{(w_1,w_2)\in R_{(3,0)}\\(w_1,w_2)\in R_{(0,3)}}}q^{\frac{7}{3}w_1+\frac{8}{3}w_2}+{\cal O}(q^{15})\rt)\nn\\
&=\frac{(1+q+q^2)(1+q+q^2+q^3+q^4)}{(q^2;q)(q^3;q)}\lt(1-q^2-q^4-q^5+q^6-q^7+q^9+{\cal O}(q^{10})\rt)\nn\\
&=1+2q+3q^2+5q^3+8q^4+11q^5+17q^6+24q^7+34q^8+47q^9+{\cal O}(q^{10}),
\ea
which agrees with the character of the $(2,2,3)$ module in the $(P=3,Q=7)$ minimal model. In summary, we obtain the following correspondence between poles and modules, 
\ba
y_1=q,\ y_2=q\;\;\quad &\leftrightarrow\quad  (1,1,5)\ {\rm vacuum\ module},\\
y_1=q^{\frac{4}{3}},\ y_2=q^{\frac{5}{3}} \quad &\leftrightarrow \quad (1,2,4)\ {\rm module},\\
y_1=q^{\frac{5}{3}},\ y_2=q^{\frac{7}{3}}\quad &\leftrightarrow \quad (1,3,3)\ {\rm module},\\
y_1=q^{\frac{7}{3}},\ y_2=q^{\frac{8}{3}}\quad &\leftrightarrow\quad (2,2,3)\ {\rm module},
\ea
which exactly reproduces the correspondence rule (\ref{eq:pole-surface}) obtained from the general discussion. 

\begin{comment}
The above poles have already reproduced all the non-degenerate modules in the $(P=3,Q=7)$ minimal model, it is then natural to conjecture that all the other poles, if the corresponding Schur index can be interpreted as a character, does not give any independent character of any new module in this model. Indeed, we can compute the Schur index associated to the pole at $x_1=q^{\frac{8}{3}}$, $x_2=q^{\frac{10}{3}}$ as
\ba
{\cal I}_{(A_2,A_3)+\mathbb{S}^{\frac{8}{3},\frac{10}{3}}}(q)&=\frac{(1+q+q^2+q^3)(1+q+q^2+q^3+q^4+q^5)}{(q^2;q)(q^3;q)}\nn\\
&\times\lt(1-q-q^3+q^4-q^5+q^6-2q^7+2q^8-q^9+{\cal O}(q^{10})\rt)\nn\\
&=1+q+2q^2+3q^3+5q^4+7q^5+11q^6+14q^7+21q^8+28q^9+{\cal O}(q^{10}),
\ea
which matches exactly with ${\cal I}_{(A_2,A_3)+\mathbb{S}^{\frac{4}{3},\frac{5}{3}}}(q)$ and thus reproduces the $(1,2,4)=(2,4,1)$ module again. 
\end{comment}

One can confirm that other poles do not give rise to any new characters, for example we have 
\ba
{\cal I}_{(A_2,A_3)}^{\mathbb{S}^{(2,4)}}(q)={\cal I}_{(A_2,A_3)}^{\mathbb{S}^{(1,2)}}(q),\quad {\cal I}_{(A_2,A_3)}^{\mathbb{S}^{(2,2)}}(q)={\cal I}_{(A_2,A_3)}^{\mathbb{S}^{(2,3)}}(q),\quad {\cal I}_{(A_2,A_3)}^{\mathbb{S}^{(1,4)}}(q)={\cal I}_{(A_2,A_3)}^{\mathbb{S}^{(1,2)}}(q),\quad \dots
\ea
The above fact reflects the permutation symmetry among $n_1\leftrightarrow n_2 \leftrightarrow n_3$ in $W_3$ minimal models. 

In the same way, we can check that the correspondence between poles and modules in $(P=3,Q=8)$ minimal model (the dual of $(A_2,A_4)$ theory) is given by\footnote{We checked this correspondence up to ${\cal O}(q^{11})$ order. We will always check this kind of correspondence up to this order throughout the paper, unless claimed otherwise.} 
\ba
y_1=q,\ y_2=q\;\;\quad &\leftrightarrow\quad  (1,1,6)\ {\rm vacuum\ module},\\
y_1=q^{\frac{5}{3}},\ y_2=q^{\frac{7}{3}}\quad &\leftrightarrow \quad (1,3,4)\ {\rm module},\\
y_1=q^{\frac{7}{3}},\ y_2=q^{\frac{8}{3}}\quad &\leftrightarrow \quad (2,3,3)\ {\rm module},\\
y_1=q^{\frac{8}{3}},\ y_2=q^{\frac{10}{3}}\quad &\leftrightarrow \quad (2,2,4)\ {\rm module}.
\ea
In the $(P=3,Q=10)$ minimal model, the correspondence is given by 
\ba
y_1=q,\ y_2=q\;\; \quad &\leftrightarrow\quad  (1,1,8)\ {\rm vacuum\ module},\\
y_1=q^{\frac{4}{3}},\ y_2=q^{\frac{5}{3}}\quad &\leftrightarrow \quad (1,2,7)\ {\rm module},\\
y_1=q^{\frac{5}{3}},\ y_2=q^{\frac{7}{3}}\quad &\leftrightarrow \quad (1,3,6)\ {\rm module},\\
y_1=q^{\frac{7}{3}},\ y_2=q^{\frac{8}{3}}\quad &\leftrightarrow \quad (2,3,5)\ {\rm module},\\
y_1=q^{\frac{8}{3}},\ y_2=q^{\frac{10}{3}}\quad &\leftrightarrow \quad (2,4,4)\ {\rm module},\\
y_1=q^{\frac{10}{3}},\ y_2=q^{\frac{11}{3}}\quad &\leftrightarrow \quad (3,3,4)\ {\rm module},\\
y_1=q^{\frac{10}{3}},\ y_2=q^{\frac{14}{3}}\quad &\leftrightarrow \quad (2,2,6)\ {\rm module},\\
y_1=q^{\frac{13}{3}},\ y_2=q^{\frac{14}{3}}\quad &\leftrightarrow \quad (1,4,5)\ {\rm module}.
\ea
The remaining poles such as $y_1=q^{\frac{11}{3}}$, $y_2=q^{\frac{13}{3}}$ give rise to the character, for example, of $(2,3,5)$ module again. 

It is clear from the above explicit computations that the pole position $y_1=q^{t_1}$ and $y_2=q^{t_2}$ in the $(A_2,A_{n-1})$ series indeed corresponds to the $(n_1,n_2,n_3)$ module in the dual $W_3$ minimal model following the rule (\ref{eq:pole-surface}), i.e. 
\ba
n_1={\rm log}_q(y_1^2y_2^{-1})=2t_1-t_2,\qquad n_2={\rm log}_q(y_2^2y_1^{-1})=2t_2-t_1,\label{rule-A2}
\ea
and $n_3$ is determined from $n_1+n_2+n_3=Q$, together with a permutation symmetry among $(n_1,n_2,n_3)$. 

We can also check the level-rank duality between $(A_1,A_2)$ theory and $(A_2,A_1)$ theory for surface operators. The Schur index of $(A_2,A_1)$ computed in the TQFT way, 
\ba
{\cal I}_{(A_2,A_1)} &= \frac{1}{(q^2;q)(q^3;q)}\sum_{(\lambda_1,\lambda_2)} \chi_{(\lambda_1,\lambda_2)}(q^\rho)f^{I_{3,-2}}_{(\lambda_1,\lambda_2)}(q^5)\nn\\
&=\frac{1}{(q^2;q)(q^3;q)}\lt(1-q^5\sum_{(w_1,w_2)\in R_{(1,1)}}q^{w_1+w_2}+q^{10}\sum_{(w_1,w_2)\in R_{(3,0)},\ R_{(0,3)}}q^{w_1+w_2}+{\cal O}(q^{15})\rt)\nn\\
&=\frac{1}{(q^2;q)(q^3;q)}\lt(1-q^3-2q^4-2q^5-2q^6-q^7+2q^7+2q^8+{\cal O}(q^{9})\rt)\nn\\
&=1+q^2+q^3+q^4+2q^5+2q^6+2q^7+3q^8+{\cal O}(q^9),
\ea
matches with the $(A_1,A_2)$ index. The pole at $y_1=q^{\frac{4}{3}}$, $y_2=q^{\frac{5}{3}}$ reproduces the character of the $(2,3)$ module of the $(P=2,Q=5)$ Lee-Yang model, or equivalently the $(1,2,2)$ module of the level-rank dual model $(P=3,Q=5)$. 
\ba
{\cal I}_{(A_2,A_1)}^{\mathbb{S}^{(1,2)}}(q)&=\frac{1+q+q^2}{(q^2;q)(q^3;q)}\lt(1-q^2-q^3-q^4-q^5+q^7+q^8+3q^9+{\cal O}(q^{10})\rt)\nn\\
&=1+q+q^2+q^3+2q^4+2q^5+3q^6+3q^7+4q^8+5q^9+{\cal O}(q^{10})\nn\\
&={\cal I}_{(A_1,A_2)}^{\mathbb{S}^{(2)}}(q):
\ea
Similarly, the index at the pole $y_1=q^{\frac{5}{3}}$, $y_2=q^{\frac{7}{3}}$, 
\ba
{\cal I}_{(A_2,A_1)}^{\mathbb{S}^{(1,3)}}(q)
&=\frac{(1+q^2)(1+q+q^2)}{(q^2;q)(q^3;q)}\lt(1-q-q^2+q^3-q^4-q^5+q^6+2q^7+q^{10}+{\cal O}(q^{11})\rt)\nn\\
&=1+q^2+q^3+q^4+q^5+2q^6+2q^7+3q^8+3q^9+4q^{10}+{\cal O}(q^{11})\nn\\
&={\cal I}_{(A_1,A_2)}(q),
\ea
corresponds to the character of the $(1,3,1)$ module, i.e. the vacuum $(1,1,3)$ module of the $(P=3,Q=5)$ minimal model.

\subsection{$(A_3,A_4)$ theory and $(A_4,A_5)$ theory
%More general minimal models $(P=N,Q)$}\label{s:general-minimal}
}

We now consider a rank-three example, the $(A_3,A_4)$ theory, and a rank-four example, the $(A_4,A_5)$ theory.

First the Schur index of the $(A_3,A_4)$ theory matches with the character of the $(1,1,1,6)$ vacuum module of $(P=4,Q=9)$ minimal model: 
\ba
{\cal I}_{(A_3,A_4)}(q)&=\frac{1}{(q^2;q)(q^3;q)(q^4;q)}\sum_{(\lambda_1,\lambda_2,\lambda_3)}\chi_{(\lambda_1,\lambda_2,\lambda_3)}(q^\rho)f^{I_{4,-3}}_{(\lambda_1,\lambda_2,\lambda_3)}(q^9)\nn\\
&=\frac{1}{(q^2;q)(q^3;q)(q^4;q)}\lt(1-q^6-2q^7-3q^8-3q^9-3q^{10}+{\cal O}(q^{11})\rt)\nn\\
&=1+q^2+2q^3+4q^4+5q^5+9q^6+12q^7+21q^8+29q^9+44q^{10}+{\cal O}(q^{11}).
\ea
It also corresponds to the Higgsed Schur index with an additional regular puncture with respect to the pole $y_1=q^{\frac{3}{2}}$, $y_2=q^2$ and $y_3=q^{\frac{3}{2}}$, i.e. $y=q^\rho$. This pole is translated according to the dictionary (\ref{eq:pole-surface}) to the trivial surface operator $\mathbb{S}^{(1,1,1)}$. 

The pole at $(y_1,y_2,y_3)=(q^{\frac{7}{4}},q^{\frac{5}{2}},q^{\frac{9}{4}})$ gives rise to the character of the $(1,1,2,5)$ module of $(P=4,Q=9)$ $W_4$ minimal model,
\ba
{\cal I}_{(A_3,A_4)}^{\mathbb{S}^{(1,1,2)}}(q)&=\frac{1+q+q^2+q^3}{(q^2;q)(q^3;q)(q^4;q)}\sum_{(\lambda_1,\lambda_2,\lambda_3)}\chi_{(\lambda_1,\lambda_2,\lambda_3)}(q^{\frac{7}{4}},q^{\frac{5}{2}},q^{\frac{9}{4}})f^{I_{4,-3}}_{(\lambda_1,\lambda_2,\lambda_3)}(q^9)\nn\\
&=\frac{1+q+q^2+q^3}{(q^2;q)(q^3;q)(q^4;q)}\lt(1-q^5-q^6-2q^7-2q^8-3q^9-2q^{10}+{\cal O}(q^{11})\rt)\nn\\
&=1+q+2q^2+4q^3+7q^4+11q^5+19q^6+28q^7+44q^8+65q^9+97q^{10}+{\cal O}(q^{11}),
\ea
as 
\ba
f_{(\frac{7}{4},\frac{5}{2},\frac{9}{4})}(q)=1+q+q^2+q^3.
\ea
\begin{comment}
We observe again that by denoting $x_i=q^{s_i}$, the label of the minimal model module is given by 
\ba
n_1=2s_1-s_2,\quad n_2=2s_2-s_1-s_3,\quad n_3=2s_3-s_2,
\ea
where the coefficients of $s_j$'s in the expression of $n_i$ are determined by the Cartan matrix $(c_{ij})$ as 
\ba
n_i=\sum_j c_{ij}s_j,\label{rule-minimal}
\ea
under the convention $c_{ii}=2$ for $^\forall i$. We conjecture that this correspondence holds for all $(A_{N-1},A_{Q-N-1})$ theories with ${\rm gcd}(N,Q)=1$. 
\end{comment}

We also checked the following correspondence in the $(A_3,A_4)$ theory, 
\ba
(y_1,y_2,y_3)=(q^2,q^3,q^3)\quad &\leftrightarrow \quad(1,1,3,4)\ {\rm module},\\
(y_1,y_2,y_3)=(q^{\frac{9}{4}},q^{\frac{7}{2}},q^{\frac{15}{4}})\quad &\leftrightarrow \quad(1,1,4,3)\ {\rm module},\\
(y_1,y_2,y_3)=(q^{\frac{9}{4}},q^{\frac{7}{2}},q^{\frac{11}{4}})\quad&\leftrightarrow \quad(1,2,2,4)\ {\rm module},\\
(y_1,y_2,y_3)=(q^{\frac{5}{2}},q^{4},q^{\frac{7}{2}})\quad &\leftrightarrow \quad (1,2,3,3)\ {\rm module},\\
(y_1,y_2,y_3)=(q^{3},q^{4},q^{3})\quad &\leftrightarrow \quad (2,2,2,3)\ {\rm module}.
\ea
It exactly matches with the translation rule (\ref{eq:pole-surface}) and that modules in the minimal model share the same label ${\bf n}=(n_1,n_2,n_3)$ with the surface operator $\mathbb{S}^{\bf n}$. 

In the $(A_4,A_5)$ theory, by using the expressions of the factor $f_{\vec{t}}(q)$, e.g. 
\ba
&f_{(\frac{11}{5},\frac{17}{5},\frac{18}{5},\frac{14}{5})}(q)=1+q+q^2+q^3+q^4,\\
&f_{(\frac{12}{5},\frac{19}{5},\frac{21}{5},\frac{18}{5})}(q)=(1+q+q^2+q^3+q^4)\frac{1-q^6}{1-q^2},
\ea
we checked the correspondence 
\ba
(y_1,y_2,y_3,y_4)=(q^2,q^3,q^3,q^2)\quad &\leftrightarrow \quad (1,1,1,1,7)\ {\rm vacuum\ module},\\
(y_1,y_2,y_3,y_4)=(q^{\frac{11}{5}},q^{\frac{17}{5}},q^{\frac{18}{5}},q^{\frac{14}{5}})\quad & \leftrightarrow \quad (1,1,1,2,6)\ {\rm module},\\
(y_1,y_2,y_3,y_4)=(q^{\frac{12}{5}},q^{\frac{19}{5}},q^{\frac{21}{5}},q^{\frac{18}{5}})\quad & \leftrightarrow \quad (1,1,1,3,5)\ {\rm module}.
\ea

\section{Generalized Euler identities}\label{a:Euler-id}

In fact the Euler identity, 
\ba
\sum_{n\in\mathbb{Z}}(-1)^n q^{\frac{3}{2}n^2\pm \frac{1}{2}n}=(q;q),
\ea
is a special case of the Jacobi triple identity, 
\ba
\sum_{n\in\mathbb{Z}}p^{n^2}z^n=\prod_{n>0}(1-p^{2n})(1+p^{2n-1}z)(1+p^{2n-1}z^{-1}),
\ea
with $p=q^{\frac{3}{2}}$ and $z=-q^{\pm \frac{1}{2}}$. 

For the Generalized Euler identity used in $G=$SU$(4)$, we can rewrite it as 
\ba
&\sum_{n_1,n_2,n_3}(-1)^{n_1+n_2}q^{\frac{3}{2}n_1^2+\frac{3}{2}n_2^2+2n_3^2+2n_1n_3+2n_2n_3+n_1n_2-\frac{1}{2}n_1-\frac{1}{2}n_2-n_3}\nn\\
&=\sum_{n_1,n_2,n_3}(-1)^{n_1+n_2}q^{(n_1+n_3)^2+(n_2+n_3)^2+\frac{1}{2}(n_1+n_2)^2-\frac{1}{2}(n_1+n_2+2n_3)}\nn\\
&=\sum_{\substack{k_1,k_2,k_3\in\mathbb{Z}\\k_1+k_2+k_3\in 2\mathbb{Z}}}(-1)^{k_3}q^{k_1^2+k_2^2+\frac{1}{2}k_3^2-\frac{1}{2}(k_1+k_2)}\nn\\
&=\sum_{\substack{k_1,k_2,k_3\in\mathbb{Z}\\k_1+k_2+k_3\in 2\mathbb{Z}}}(-1)^{k_1+k_2}q^{k_1^2+k_2^2+\frac{1}{2}k_3^2-\frac{1}{2}(k_1+k_2)},
\ea
where we set $k_1=n_1+n_3$, $k_2=n_2+n_3$ and $k_3=n_1+n_2$. Using the equality in the last line, we further have 
\ba
\sum_{\substack{k_1,k_2,k_3\in\mathbb{Z}\\k_1+k_2+k_3\in 2\mathbb{Z}}}(-1)^{k_3}q^{k_1^2+k_2^2+\frac{1}{2}k_3^2-\frac{1}{2}(k_1+k_2)}=\sum_{k_1,k_2,k_3\in\mathbb{Z}}\frac{1+(-1)^{k_1+k_2+k_3}}{2}(-1)^{k_3}q^{k_1^2+k_2^2+\frac{1}{2}k_3^2-\frac{1}{2}(k_1+k_2)}.
%=\sum_{k_1,k_2,k_3\in\mathbb{Z}}(-1)^{k_3}q^{k_1^2+k_2^2+\frac{1}{2}k_3^2-\frac{1}{2}(k_1+k_2)}.
\ea
Substituting $(p,z)=(q,\pm q^{-\frac{1}{2}}),\ (q^{\frac{1}{2}},\mp 1)$ into the Jacobi triple identity, we obtain 
\ba
\sum_{k\in\mathbb{Z}}q^{k^2-\frac{1}{2}k}=\prod_{n>0}(1-q^{2n})(1+q^{n-\frac{1}{2}}),\\
\sum_{k\in\mathbb{Z}}(-1)^kq^{\frac{1}{2}k^2}=\prod_{n>0}(1-q^n)(1-q^{n-\frac{1}{2}})^2,\\
\sum_{k\in\mathbb{Z}}(-1)^kq^{k^2-\frac{1}{2}k}=\prod_{n>0}(1-q^{2n})(1-q^{n-\frac{1}{2}}),\\
\sum_{k\in\mathbb{Z}}q^{\frac{1}{2}k^2}=\prod_{n>0}(1-q^n)(1+q^{n-\frac{1}{2}})^2,
\ea
and therefore 
\ba
&\sum_{n_1,n_2,n_3}(-1)^{n_1+n_2}q^{\frac{3}{2}n_1^2+\frac{3}{2}n_2^2+2n_3^2+2n_1n_3+2n_2n_3-n_1n_2-\frac{1}{2}n_1-\frac{1}{2}n_2-n_3}\nn\\
&=\sum_{n_1,n_2,n_3}(-1)^{n_3}q^{\frac{3}{2}n_1^2+\frac{3}{2}n_2^2+2n_3^2+2n_1n_3+2n_2n_3-n_1n_2-\frac{1}{2}n_1-\frac{1}{2}n_2-n_3}\nn\\
&=\prod_{n>0}(1-q^n)(1-q^{2n})^2(1-q^{2n-1})^2=(q;q)^3.
\ea

For $G=$SU$(5)$, we conjecture that the following generalized Euler identity holds, 
\ba
(q;q)^6=\sum_{n_{1,2,3,4,5,6}\in\mathbb{Z}}&(-1)^{n_1+n_2+n_3+n_6}q^{\frac{3}{2}n_1^2+\frac{3}{2}n_2^2+\frac{3}{2}n_3^2+2n_4^2+2n_5^2+\frac{5}{2}n_{6}^2}\nn\\
&\times q^{n_1n_2+2n_1n_4+n_1n5+2n_1n_6+n_2n_3+2n_2n_4+2n_2n_5+2n_2n_6}\nn\\
&\times q^{n_3n_4+2n_3n_5+2n_3n_6+2n_4n_5+3n_4n_6+3n_5n_6}\nn\\
&\times q^{-\frac{1}{2}n_1-\frac{1}{2}n_2-\frac{1}{2}n_3-n_4-n_5-\frac{3}{2}n_6}.
\ea

\bibliography{surface-operator}

\providecommand{\href}[2]{#2}\begingroup\raggedright\begin{thebibliography}{10}

\bibitem{AGT}
L.~F. Alday, D.~Gaiotto, and Y.~Tachikawa, ``Liouville Correlation Functions
  from Four-dimensional Gauge Theories,''
  \href{http://dx.doi.org/10.1007/s11005-010-0369-5}{{\em Lett.Math.Phys.}
  {\bfseries 91} (2010) 167--197},
\href{http://arxiv.org/abs/0906.3219}{{\ttfamily arXiv:0906.3219 [hep-th]}}.
%%CITATION = ARXIV:0906.3219;%%.

\bibitem{Wyllard:2009hg}
N.~Wyllard, ``{A(N-1) Conformal Toda Field Theory Correlation Functions from
  Conformal ${\mathcal{N}}\!=2$ $SU(N)$ Quiver Gauge Theories},''
  \href{http://dx.doi.org/10.1088/1126-6708/2009/11/002}{{\em JHEP} {\bfseries
  11} (2009) 002},
\href{http://arxiv.org/abs/0907.2189}{{\ttfamily arXiv:0907.2189 [hep-th]}}.
%%CITATION = ARXIV:0907.2189;%%.

\bibitem{GRRY}
A.~Gadde, L.~Rastelli, S.~S. Razamat, and W.~Yan, ``{The 4d Superconformal
  Index from q-deformed 2d Yang-Mills},'' {\em Phys.Rev.Lett.} {\bfseries 106}
  (2011) 241602,
\href{http://arxiv.org/abs/1104.3850}{{\ttfamily arXiv:1104.3850 [hep-th]}}.
%%CITATION = ARXIV:1104.3850;%%.

\bibitem{Gadde:2011uv}
A.~Gadde, L.~Rastelli, S.~S. Razamat, and W.~Yan, ``{Gauge Theories and
  Macdonald Polynomials},''
  \href{http://dx.doi.org/10.1007/s00220-012-1607-8}{{\em Commun. Math. Phys.}
  {\bfseries 319} (2013) 147--193},
\href{http://arxiv.org/abs/1110.3740}{{\ttfamily arXiv:1110.3740 [hep-th]}}.
%%CITATION = ARXIV:1110.3740;%%.

\bibitem{classS}
D.~Gaiotto, ``{N=2 dualities},'' {\em JHEP} {\bfseries 08} (2012) 034,
\href{http://arxiv.org/abs/0904.2715}{{\ttfamily arXiv:0904.2715 [hep-th]}}.
%\%CITATION = ARXIV:0904.2715;\%\%.

\bibitem{Gaiotto:2009hg}
D.~Gaiotto, G.~W. Moore, and A.~Neitzke, ``{Wall-Crossing, Hitchin Systems, and
  the Wkb Approximation},''
\href{http://arxiv.org/abs/0907.3987}{{\ttfamily arXiv:0907.3987 [hep-th]}}.
%%CITATION = ARXIV:0907.3987;%%.

\bibitem{Beem:2016wfs}
C.~Beem, L.~Rastelli, and B.~C. van Rees, ``{More ${\mathcal N}=4$
  superconformal bootstrap},''
  \href{http://dx.doi.org/10.1103/PhysRevD.96.046014}{{\em Phys. Rev.}
  {\bfseries D96} no.~4, (2017) 046014},
\href{http://arxiv.org/abs/1612.02363}{{\ttfamily arXiv:1612.02363 [hep-th]}}.
%%CITATION = ARXIV:1612.02363;%%.

\bibitem{Gadde:2009kb}
A.~Gadde, E.~Pomoni, L.~Rastelli, and S.~S. Razamat, ``{S-Duality and 2D
  Topological QFT},'' \href{http://dx.doi.org/10.1007/JHEP03(2010)032}{{\em
  JHEP} {\bfseries 03} (2010) 032},
\href{http://arxiv.org/abs/0910.2225}{{\ttfamily arXiv:0910.2225 [hep-th]}}.
%%CITATION = ARXIV:0910.2225;%%.

\bibitem{Buican:2014hfa}
M.~Buican, S.~Giacomelli, T.~Nishinaka, and C.~Papageorgakis,
  ``{Argyres-Douglas Theories and S-Duality},''
  \href{http://dx.doi.org/10.1007/JHEP02(2015)185}{{\em JHEP} {\bfseries 02}
  (2015) 185},
\href{http://arxiv.org/abs/1411.6026}{{\ttfamily arXiv:1411.6026 [hep-th]}}.
%%CITATION = ARXIV:1411.6026;%%.

\bibitem{Xie:2016uqq}
D.~Xie and S.-T. Yau, ``{New ${\mathcal{N}}\!=2$ Dualities},''
\href{http://arxiv.org/abs/1602.03529}{{\ttfamily arXiv:1602.03529 [hep-th]}}.
%%CITATION = ARXIV:1602.03529;%%.

\bibitem{Xie:2017vaf}
D.~Xie and S.-T. Yau, ``{Argyres-Douglas Matter and ${\mathcal{N}}\!=2$
  Dualities},''
\href{http://arxiv.org/abs/1701.01123}{{\ttfamily arXiv:1701.01123 [hep-th]}}.
%%CITATION = ARXIV:1701.01123;%%.

\bibitem{Buican:2017fiq}
M.~Buican, Z.~Laczko, and T.~Nishinaka, ``{$ \mathcal{N} $ = 2 S-duality
  revisited},'' \href{http://dx.doi.org/10.1007/JHEP09(2017)087}{{\em JHEP}
  {\bfseries 09} (2017) 087},
\href{http://arxiv.org/abs/1706.03797}{{\ttfamily arXiv:1706.03797 [hep-th]}}.
%%CITATION = ARXIV:1706.03797;%%.

\bibitem{Xie:2017aqx}
D.~Xie and K.~Ye, ``{Argyres-Douglas Matter and S-Duality: Part II},''
  \href{http://dx.doi.org/10.1007/JHEP03(2018)186}{{\em JHEP} {\bfseries 03}
  (2018) 186},
\href{http://arxiv.org/abs/1711.06684}{{\ttfamily arXiv:1711.06684 [hep-th]}}.
%%CITATION = ARXIV:1711.06684;%%.

\bibitem{Buican:2016arp}
M.~Buican and T.~Nishinaka, ``{Conformal Manifolds in Four Dimensions and
  Chiral Algebras},''
  \href{http://dx.doi.org/10.1088/1751-8113/49/46/465401}{{\em J. Phys.}
  {\bfseries A49} no.~46, (2016) 465401},
\href{http://arxiv.org/abs/1603.00887}{{\ttfamily arXiv:1603.00887 [hep-th]}}.
%%CITATION = ARXIV:1603.00887;%%.

\bibitem{chiralsymm}
C.~Beem, M.~Lemos, P.~Liendo, W.~Peelaers, L.~Rastelli, and B.~C. van Rees,
  ``{Infinite Chiral Symmetry in Four Dimensions},'' {\em Comm. Math. Phys.}
  {\bfseries 336} no.~3, (2015) 1359--1433,
\href{http://arxiv.org/abs/1312.5344}{{\ttfamily arXiv:1312.5344 [hep-th]}}.
%%CITATION = ARXIV:1312.5344v3;%%.

\bibitem{Buican:2017rya}
M.~Buican and Z.~Laczko, ``{Nonunitary Lagrangians and Unitary Non-Lagrangian
  Conformal Field Theories},''
  \href{http://dx.doi.org/10.1103/PhysRevLett.120.081601}{{\em Phys. Rev.
  Lett.} {\bfseries 120} no.~8, (2018) 081601},
\href{http://arxiv.org/abs/1711.09949}{{\ttfamily arXiv:1711.09949 [hep-th]}}.
%%CITATION = ARXIV:1711.09949;%%.

\bibitem{CGS-17-1}
C.~Cordova, D.~Gaiotto, and S.-H. Shao, ``{Surface Defect Indices and 2d-4d BPS
  States},'' {\em JHEP} {\bfseries 12} (2017) 078,
\href{http://arxiv.org/abs/1703.02525}{{\ttfamily arXiv:1703.02525 [hep-th]}}.
%\%CITATION = ARXIV:1703.02525;\%\%.

\bibitem{CGS-17-2}
C.~Cordova, D.~Gaiotto, and S.-H. Shao, ``{Surface Defects and Chiral
  Algebras},'' {\em JHEP} {\bfseries 05} (2017) 140,
\href{http://arxiv.org/abs/1704.01955}{{\ttfamily arXiv:1704.01955 [hep-th]}}.
%\%CITATION = ARXIV:1704.01955;\%\%.

\bibitem{CGS-line}
C.~Cordova, D.~Gaiotto, and S.-H. Shao, ``{Infrared Computations of Defect
  Schur Indices},'' {\em JHEP} {\bfseries 11} (2016) 106,
\href{http://arxiv.org/abs/1606.08429}{{\ttfamily arXiv:1606.08429 [hep-th]}}.
%\%CITATION = ARXIV:1606.08429v2;\%\%.

\bibitem{Argyres:1995jj}
P.~C. Argyres and M.~R. Douglas, ``{New Phenomena in $SU(3)$ Supersymmetric
  Gauge Theory},'' \href{http://dx.doi.org/10.1016/0550-3213(95)00281-V}{{\em
  Nucl. Phys.} {\bfseries B448} (1995) 93--126},
\href{http://arxiv.org/abs/hep-th/9505062}{{\ttfamily arXiv:hep-th/9505062
  [hep-th]}}.
%%CITATION = HEP-TH/9505062;%%.

\bibitem{Argyres:1995xn}
P.~C. Argyres, M.~R. Plesser, N.~Seiberg, and E.~Witten, ``{New
  ${\mathcal{N}}\!=2$ Superconformal Field Theories in Four-Dimensions},''
  \href{http://dx.doi.org/10.1016/0550-3213(95)00671-0}{{\em Nucl. Phys.}
  {\bfseries B461} (1996) 71--84},
\href{http://arxiv.org/abs/hep-th/9511154}{{\ttfamily arXiv:hep-th/9511154
  [hep-th]}}.
%%CITATION = HEP-TH/9511154;%%.

\bibitem{Eguchi:1996vu}
T.~Eguchi, K.~Hori, K.~Ito, and S.-K. Yang, ``{Study of ${\mathcal{N}}\!=2$
  Superconformal Field Theories in Four-Dimensions},''
  \href{http://dx.doi.org/10.1016/0550-3213(96)00188-5}{{\em Nucl. Phys.}
  {\bfseries B471} (1996) 430--444},
\href{http://arxiv.org/abs/hep-th/9603002}{{\ttfamily arXiv:hep-th/9603002
  [hep-th]}}.
%%CITATION = HEP-TH/9603002;%%.

\bibitem{Cecotti:2010fi}
S.~Cecotti, A.~Neitzke, and C.~Vafa, ``{R-Twisting and 4D/2D
  Correspondences},''
\href{http://arxiv.org/abs/1006.3435}{{\ttfamily arXiv:1006.3435 [hep-th]}}.
%%CITATION = ARXIV:1006.3435;%%.

\bibitem{Bonelli:2011aa}
G.~Bonelli, K.~Maruyoshi, and A.~Tanzini, ``{Wild Quiver Gauge Theories},''
  \href{http://dx.doi.org/10.1007/JHEP02(2012)031}{{\em JHEP} {\bfseries 02}
  (2012) 031},
\href{http://arxiv.org/abs/1112.1691}{{\ttfamily arXiv:1112.1691 [hep-th]}}.
%%CITATION = ARXIV:1112.1691;%%.

\bibitem{Xie:2012hs}
D.~Xie, ``{General Argyres-Douglas Theory},''
  \href{http://dx.doi.org/10.1007/JHEP01(2013)100}{{\em JHEP} {\bfseries 01}
  (2013) 100},
\href{http://arxiv.org/abs/1204.2270}{{\ttfamily arXiv:1204.2270 [hep-th]}}.
%%CITATION = ARXIV:1204.2270;%%.

\bibitem{Cordova-Shao}
C.~Cordova and S.-H. Shao, ``{Schur Indices, BPS Particles, and Argyres-Douglas
  Theories},'' {\em JHEP} {\bfseries 01} (2016) 040,
\href{http://arxiv.org/abs/1506.00265}{{\ttfamily arXiv:1506.00265 [hep-th]}}.
%\%CITATION = ARXIV:1506.00265v2;\%\%.

\bibitem{Ito:2017ypt}
K.~Ito and H.~Shu, ``{ODE/IM Correspondence and the Argyres-Douglas Theory},''
  \href{http://dx.doi.org/10.1007/JHEP08(2017)071}{{\em JHEP} {\bfseries 08}
  (2017) 071},
\href{http://arxiv.org/abs/1707.03596}{{\ttfamily arXiv:1707.03596 [hep-th]}}.
%%CITATION = ARXIV:1707.03596;%%.

\bibitem{Ito:2018ypt}
K.~Ito, M.~Mari$\tilde{{\rm n}}$o, and H.~Shu, ``{TBA equations and resurgent
  Quantum Mechanics},'' {\em arXiv preprint} (2018) ,
\href{http://arxiv.org/abs/1811.04812}{{\ttfamily arXiv:1811.04812 [hep-th]}}.
%%CITATION = ARXIV:1811.04812v2;%%.

\bibitem{Creutzig-17}
T.~Creutzig, ``{W-algebras for Argyres-Douglas theories},'' {\em ArXiv
  e-prints} (2017) ,
\href{http://arxiv.org/abs/1701.05926}{{\ttfamily arXiv:1701.05926 [hep-th]}}.
%\%CITATION = ARXIV:1701.05926;\%\%.

\bibitem{Beem:2017ooy}
C.~Beem and L.~Rastelli, ``{Vertex Operator Algebras, Higgs Branches, and
  Modular Differential Equations},''
  \href{http://dx.doi.org/10.1007/JHEP08(2018)114}{{\em JHEP} {\bfseries 08}
  (2018) 114},
\href{http://arxiv.org/abs/1707.07679}{{\ttfamily arXiv:1707.07679 [hep-th]}}.
%%CITATION = ARXIV:1707.07679;%%.

\bibitem{Creutzig-18}
T.~Creutzig, ``{Logarithmic W-algebras and Argyres-Douglas theories at higher
  rank},'' {\em ArXiv e-prints} (2018) ,
\href{http://arxiv.org/abs/1809.01725}{{\ttfamily arXiv:1809.01725 [hep-th]}}.
%\%CITATION = ARXIV:1809.01725;\%\%.

\bibitem{Choi-Nishinaka}
J.~Choi and T.~Nishinaka, ``{On the chiral algebra of Argyres-Douglas theories
  and S-duality},'' {\em JHEP} {\bfseries 04} (2018) 004,
\href{http://arxiv.org/abs/1711.07941}{{\ttfamily arXiv:1711.07941 [hep-th]}}.
%\%CITATION = ARXIV:1711.07941v2;\%\%.

\bibitem{Buican-Nishinaka15}
M.~Buican and T.~Nishinaka, ``{On the superconformal index of Argyres--Douglas
  theories},'' {\em Journal of Physics A: Mathematical and Theoretical}
  {\bfseries 49} no.~1, (2015) 015401,
\href{http://arxiv.org/abs/1505.05884}{{\ttfamily arXiv:1505.05884 [hep-th]}}.
%\%CITATION = ARXIV:1505.05884v2;\%\%.

\bibitem{Buican-Nishinaka17}
M.~Buican and T.~Nishinaka, ``{On Irregular Singularity Wave Functions and
  Superconformal Indices},'' {\em JHEP} {\bfseries 09} (2017) 066,
\href{http://arxiv.org/abs/1705.07173}{{\ttfamily arXiv:1705.07173 [hep-th]}}.
%\%CITATION = ARXIV:1705.07173;\%\%.

\bibitem{triplet-W}
B.~Feigin, A.~M. Gainutdinov, A.~M. Semikhatov, and I.~Y. Tipunin,
  ``{Kazhdan-Lusztig correspondence for the representation category of the
  triplet W-algebra in logarithmic CFT},'' {\em Theor. Math. Phys.} {\bfseries
  148} (2006) 1210--1235.

\bibitem{logrithmic-minimal}
B.~L. Feigin, A.~M. Gainutdinov, A.~M. Semikhatov, and I.~Y. Tipunin,
  ``{Logarithmic extensions of minimal models: characters and modular
  transformations},'' {\em Nucl.Phys.B} {\bfseries 757} (2006) 303--343,
\href{http://arxiv.org/abs/hep-th/0606196}{{\ttfamily arXiv:hep-th/0606196
  [hep-th]}}.
%\%CITATION = ARXIV:hep-th/0606196v3;\%\%.

\bibitem{AM}
D.~Adamovic and A.~Milas, ``{On the triplet vertex algebra W(p)},'' {\em Adv.
  Math.} {\bfseries 217} (2008) 2664--2699,
\href{http://arxiv.org/abs/0707.1857}{{\ttfamily arXiv:0707.1857 [hep-th]}}.
%\%CITATION = ARXIV:0707.1857;\%\%.

\bibitem{Tsuchiya-Wood}
A.~Tsuchiya and S.~Wood, ``{The tensor structure on the representation category
  of the $W_p$ triplet algebra},'' {\em J. Phys.} {\bfseries A46} (2013)
  445203,
\href{http://arxiv.org/abs/1201.0419}{{\ttfamily arXiv:1201.0419 [hep-th]}}.
%\%CITATION = ARXIV:1201.0419;\%\%.

\bibitem{CRW}
T.~Creutzig, D.~Ridout, and S.~Wood, ``{Coset Constructions of Logarithmic (1,
  P) Models},'' \href{http://dx.doi.org/10.1007/s11005-014-0680-7}{{\em Lett.
  Math. Phys.} {\bfseries 104} (2014) 553--583},
\href{http://arxiv.org/abs/1305.2665}{{\ttfamily arXiv:1305.2665 [math.QA]}}.
%%CITATION = ARXIV:1305.2665;%%.

\bibitem{Buican:2015tda}
M.~Buican and T.~Nishinaka, ``{Argyres-Douglas Theories, the Macdonald Index,
  and an RG Inequality},''
  \href{http://dx.doi.org/10.1007/JHEP02(2016)159}{{\em JHEP} {\bfseries 02}
  (2016) 159},
\href{http://arxiv.org/abs/1509.05402}{{\ttfamily arXiv:1509.05402 [hep-th]}}.
%%CITATION = ARXIV:1509.05402;%%.

\bibitem{Buican:2015hsa}
M.~Buican and T.~Nishinaka, ``{Argyres--Douglas theories, S$^1$ reductions, and
  topological symmetries},''
  \href{http://dx.doi.org/10.1088/1751-8113/49/4/045401}{{\em J. Phys.}
  {\bfseries A49} no.~4, (2016) 045401},
\href{http://arxiv.org/abs/1505.06205}{{\ttfamily arXiv:1505.06205 [hep-th]}}.
%%CITATION = ARXIV:1505.06205;%%.

\bibitem{GRR-Higgs}
D.~Gaiotto, L.~Rastelli, and S.~S. Razamat, ``{Bootstrapping the superconformal
  index with surface defects},'' {\em JHEP} {\bfseries 01} (2013) 022,
\href{http://arxiv.org/abs/1207.3577}{{\ttfamily arXiv:1207.3577 [hep-th]}}.
%\%CITATION = ARXIV:1207.3577;\%\%.

\bibitem{Song-TQFT}
J.~Song, ``{Superconformal indices of generalized Argyres-Douglas theories from
  2d TQFT},'' {\em JHEP} {\bfseries 02} (2016) 045,
\href{http://arxiv.org/abs/1509.06730}{{\ttfamily arXiv:1509.06730 [hep-th]}}.
%\%CITATION = ARXIV:1509.06730;\%\%.

\bibitem{Alday:2013kda}
L.~F. Alday, M.~Bullimore, M.~Fluder, and L.~Hollands, ``{Surface Defects, the
  Superconformal Index and Q-Deformed Yang-Mills},''
  \href{http://dx.doi.org/10.1007/JHEP10(2013)018}{{\em JHEP} {\bfseries 10}
  (2013) 018},
\href{http://arxiv.org/abs/1303.4460}{{\ttfamily arXiv:1303.4460 [hep-th]}}.
%%CITATION = ARXIV:1303.4460;%%.

\bibitem{Bullimore:2014nla}
M.~Bullimore, M.~Fluder, L.~Hollands, and P.~Richmond, ``{The Superconformal
  Index and an Elliptic Algebra of Surface Defects},''
  \href{http://dx.doi.org/10.1007/JHEP10(2014)062}{{\em JHEP} {\bfseries 10}
  (2014) 062},
\href{http://arxiv.org/abs/1401.3379}{{\ttfamily arXiv:1401.3379 [hep-th]}}.
%%CITATION = ARXIV:1401.3379;%%.

\bibitem{Chen:2014rca}
H.-Y. Chen and H.-Y. Chen, ``{Heterotic Surface Defects and Dualities from
  2D/4D Indices},'' \href{http://dx.doi.org/10.1007/JHEP10(2014)004}{{\em JHEP}
  {\bfseries 10} (2014) 004},
\href{http://arxiv.org/abs/1407.4587}{{\ttfamily arXiv:1407.4587 [hep-th]}}.
%%CITATION = ARXIV:1407.4587;%%.

\bibitem{Maruyoshi:2016caf}
K.~Maruyoshi and J.~Yagi, ``{Surface Defects as Transfer Matrices},''
  \href{http://dx.doi.org/10.1093/ptep/pt$W_1$51}{{\em PTEP} {\bfseries 2016}
  no.~11, (2016) 113B01},
\href{http://arxiv.org/abs/1606.01041}{{\ttfamily arXiv:1606.01041 [hep-th]}}.
%%CITATION = ARXIV:1606.01041;%%.

\bibitem{Ito:2016fpl}
Y.~Ito and Y.~Yoshida, ``{Superconformal index with surface defects for class
  ${\cal S}_k$},''
\href{http://arxiv.org/abs/1606.01653}{{\ttfamily arXiv:1606.01653 [hep-th]}}.
%%CITATION = ARXIV:1606.01653;%%.

\bibitem{Nazzal:2018brc}
B.~Nazzal and S.~S. Razamat, ``{Surface Defects in E-String Compactifications
  and the Van Diejen Model},''
  \href{http://dx.doi.org/10.3842/SIGMA.2018.036}{{\em SIGMA} {\bfseries 14}
  (2018) 036},
\href{http://arxiv.org/abs/1801.00960}{{\ttfamily arXiv:1801.00960 [hep-th]}}.
%%CITATION = ARXIV:1801.00960;%%.

\bibitem{Razamat:2018zel}
S.~S. Razamat, ``{Flavored surface defects in $4d\ \mathcal N=1$ SCFTs},''
\href{http://arxiv.org/abs/1808.09509}{{\ttfamily arXiv:1808.09509 [hep-th]}}.
%%CITATION = ARXIV:1808.09509;%%.

\bibitem{Buican:2018ddk}
M.~Buican, Z.~Laczko, and T.~Nishinaka, ``{Flowing from 16 to 32
  Supercharges},'' \href{http://dx.doi.org/10.1007/JHEP10(2018)175}{{\em JHEP}
  {\bfseries 10} (2018) 175},
\href{http://arxiv.org/abs/1807.02785}{{\ttfamily arXiv:1807.02785 [hep-th]}}.
%%CITATION = ARXIV:1807.02785;%%.

\bibitem{Beem:2014rza}
C.~Beem, W.~Peelaers, L.~Rastelli, and B.~C. van Rees, ``{Chiral Algebras of
  Class S},'' \href{http://dx.doi.org/10.1007/JHEP05(2015)020}{{\em JHEP}
  {\bfseries 05} (2015) 020},
\href{http://arxiv.org/abs/1408.6522}{{\ttfamily arXiv:1408.6522 [hep-th]}}.
%%CITATION = ARXIV:1408.6522;%%.

\bibitem{Fateev-Lukyanov}
V.~A. Fateev and S.~Lukyanov, ``The Models of Two-Dimensional Conformal Quantum
  Field Theory with Z(n) Symmetry,'' {\em Int. J. Mod. Phys. A} {\bfseries 3}
  (1988) 507.

\bibitem{level-rank}
D.~Altschuler, M.~Bauer, and H.~Saleur, ``Level-rank duality in nonunitary
  coset theories,'' {\em Journal of Physics A: Mathematical and General}
  {\bfseries 23} no.~16, (1990) 789--793.

\bibitem{Song:2017oew}
J.~Song, D.~Xie, and W.~Yan, ``{Vertex Operator Algebras of Argyres-Douglas
  Theories from M5-Branes},''
  \href{http://dx.doi.org/10.1007/JHEP12(2017)123}{{\em JHEP} {\bfseries 12}
  (2017) 123},
\href{http://arxiv.org/abs/1706.01607}{{\ttfamily arXiv:1706.01607 [hep-th]}}.
%%CITATION = ARXIV:1706.01607;%%.

\bibitem{Feigin:2004wb}
B.~L. Feigin and A.~M. Semikhatov, ``{W(2)(N) Algebras},''
  \href{http://dx.doi.org/10.1016/j.nuclphysb.2004.06.056}{{\em Nucl. Phys.}
  {\bfseries B698} (2004) 409--449},
\href{http://arxiv.org/abs/math/0401164}{{\ttfamily arXiv:math/0401164
  [math-qa]}}.
%%CITATION = MATH/0401164;%%.

\bibitem{Creutzig:2017uxh}
T.~Creutzig and D.~Gaiotto, ``{Vertex Algebras for S-Duality},''
\href{http://arxiv.org/abs/1708.00875}{{\ttfamily arXiv:1708.00875 [hep-th]}}.
%%CITATION = ARXIV:1708.00875;%%.

\bibitem{Gaiotto:2017euk}
D.~Gaiotto and M.~Rap{\v c}{\'a}k, ``{Vertex Algebras at the Corner},''
\href{http://arxiv.org/abs/1703.00982}{{\ttfamily arXiv:1703.00982 [hep-th]}}.
%%CITATION = ARXIV:1703.00982;%%.

\bibitem{Song-Macdonald}
J.~Song, ``{Macdonald Index and Chiral Algebra},'' {\em JHEP} {\bfseries 08}
  (2017) 044,
\href{http://arxiv.org/abs/1612.08956}{{\ttfamily arXiv:1612.08956 [hep-th]}}.
%\%CITATION = ARXIV:1612.08956v3;\%\%.

\bibitem{Bonetti:2018fqz}
F.~Bonetti, C.~Meneghelli, and L.~Rastelli, ``{VOAs labelled by complex
  reflection groups and $4d$ SCFTs},''
\href{http://arxiv.org/abs/1810.03612}{{\ttfamily arXiv:1810.03612 [hep-th]}}.
%%CITATION = ARXIV:1810.03612;%%.

\bibitem{FNMZ}
M.~Fukuda, S.~Nakamura, Y.~Matsuo, and R.-D. Zhu, ``{SH$^c$ Realization of
  Minimal Model CFT: Triality, Poset and Burge Condition},'' {\em JHEP}
  {\bfseries 11} (2015) 168,
\href{http://arxiv.org/abs/1509.01000}{{\ttfamily arXiv:1509.01000 [hep-th]}}.
%\%CITATION = ARXIV:1509.01000;\%\%.

\bibitem{BFS-burge}
V.~Belavin, O.~Foda, and R.~Santachiara, ``{AGT, N-Burge partitions and $W_N$
  minimal models},'' {\em JHEP} {\bfseries 10} (2015) 073,
\href{http://arxiv.org/abs/1507.03540}{{\ttfamily arXiv:1507.03540 [hep-th]}}.
%\%CITATION = ARXIV:1507.03540v2;\%\%.

\bibitem{Gaiotto-Kim}
D.~Gaiotto and H.-C. Kim, ``{Surface defects and instanton partition
  functions},'' {\em arXiv preprint} (2014) ,
\href{http://arxiv.org/abs/1412.2781}{{\ttfamily arXiv:1412.2781 [hep-th]}}.
%%CITATION = ARXIV:1412.2781;%%.

\bibitem{Gukov-Witten1}
S.~Gukov and E.~Witten, ``{Gauge Theory, Ramification, And The Geometric
  Langlands Program},'' {\em arXiv preprint} (2006) ,
\href{http://arxiv.org/abs/hep-th/0612073}{{\ttfamily arXiv:hep-th/0612073
  [hep-th]}}.
%%CITATION = ARXIV:hep-th/0612073;%%.

\bibitem{Gukov-Witten2}
S.~Gukov and E.~Witten, ``{Rigid Surface Operators},'' {\em arXiv preprint}
  (2008) ,
\href{http://arxiv.org/abs/0804.1561}{{\ttfamily arXiv:0804.1561 [hep-th]}}.
%%CITATION = ARXIV:0804.1561;%%.

\bibitem{Nilpotent-defect}
O.~Chacaltana, J.~Distler, and Y.~Tachikawa, ``{Nilpotent Orbits and
  Codimension-2 Defects of 6d ${\cal N}=(2,0)$ Theories},'' {\em Modern Physics
  A} {\bfseries 28} no.~03n04, (2013) 1340006,
\href{http://arxiv.org/abs/1203.2930}{{\ttfamily arXiv:1203.2930 [hep-th]}}.
%\%CITATION = ARXIV:1203.2930;\%\%.

\end{thebibliography}\endgroup

\end{document}